\documentclass[12pt]{revtex4}

\usepackage{amsmath,amsfonts}
\usepackage{amssymb,latexsym}
\usepackage{color}
\usepackage{graphicx}
\usepackage{pdfpages}
\usepackage[subrefformat=parens]{subcaption}
\usepackage{ulem}
\usepackage{verbatim}

\usepackage{color}
\usepackage{xcolor}
\usepackage[breaklinks,colorlinks]{hyperref}
\hypersetup{
    colorlinks, 
    linktoc=all, 
    linkcolor=red,  
  citecolor=hyptxt,
  urlcolor=blue}
  \hypersetup{
  citebordercolor=,
  filebordercolor=green,
  linkbordercolor=blue
}
\definecolor{hyptxt}{rgb}{0.7, 0.4, 0.9}

\def\R{\mathbb{R}}

\def\lg{\langle }
\def\rg{\rangle }

\def\ud{\mathrm{d}}
\def\ii{\mathrm{i}}
\def\erfc{\rm Erfc}
\def\cab{\chi_{(a,b)}}
\def\hq{\hat q}
\def\hp{\hat p}

\DeclareRobustCommand{\frcshape}{\fontfamily{frc}\selectfont}
\DeclareTextFontCommand{\textfrc}{\frcshape}

\begin{document}

\title{Quantum  Smooth Boundary Forces from Constrained Geometries }
\author{J.-P. Gazeau} 
\affiliation{APC, Univ Paris Diderot, Sorbonne Paris Cit\'e -  75205 Paris, France}
\affiliation{Centro Brasileiro de Pesquisas F\'{i}sicas - 22290-180 - Rio de Janeiro, RJ, Brazil }
\author{T. Koide}
\affiliation{ Instituto de F\'{i}sica, Universidade Federal do Rio de Janeiro -  C.P.
68528, 21941-972, Rio de Janeiro, RJ, Brazil}
\author{D. Noguera}
\affiliation{Centro Brasileiro de Pesquisas F\'{i}sicas,  Rua Xavier Sigaud 150, 22290-180, Rio de Janeiro, RJ, Brazil}

\begin{abstract}
We implement the so-called Weyl-Heisenberg covariant integral quantization in the case of a classical system constrained by a bounded or semi-bounded geometry. 
The procedure, which is free of the ordering problem of operators, is illustrated with the basic example of the one-dimensional motion of a free particle in an interval, and yields 
a fuzzy boundary, a position-dependent mass (PDM), and an extra potential on the quantum level.
The consistency of our quantization is discussed by analyzing 
the semi-classical phase space portrait of the derived quantum dynamics, which is obtained as a regularization of its original classical counterpart.
\end{abstract}

\maketitle

\tableofcontents

\section{Introduction}

In this paper, we study, through a well-known textbook example, the quantization of a one-dimensional classical system which is constrained to lie within a bounded or semi-bounded set in the line. As is well-known, the canonical quantization itself is not straightforwardly applicable to non-trivial geometries, see for instance \cite{costa81,schujaff03} and references therein. The aim of our study is to propose a new approach in considering these questions. 

Our approach is an adaptation of the so-called 
Weyl-Heisenberg covariant integral quantization \cite{bergaz14,becugaro17,gazeau18}. 
For the sake of simplicity, we consider the standard coherent state (CS) quantization 
among various integral quantizations built from positive operator-valued measures (POVM) and generalize it to apply to the systems with geometric constraints.
Our quantization procedure smooths the discontinuous classical bounded or semi-bounded geometry. It 
also leads to a position-dependent mass  (PDM) and a potential term which cannot be deleted through a modification of  the kinetic operator in the quantum Hamiltonian with PDM.

Note that a seemingly similar study was carried out by one of the present authors in \cite{bafrega15},
where an exploration of the quantization of constraints in the plane is presented when standard coherent states and more general operators like thermal states  are used as ``quantizers''. 
Then two different  approaches were considered to be applied to simple examples.
The first one consists in considering an implicit constraint determined from a distribution, like a Dirac delta, viewed as a classical observable. 
The second one follows the Dirac procedure \cite{dirac64} for constraint quantization. 
A semi-classical analysis was then carried out through lower symbols and their generalizations.
In the present work, we implement a method similar to the  approach developed in \cite{bafrega15}. 
However,  contrarily  to \cite{bafrega15}, the geometric constraint is restricted to the configuration space and not to the whole phase space. Furthermore, we proceed with a comprehensive analysis of the physical consequences of such geometric constraints on the quantum observables and their semi-classical portraits established through the use of  Gaussian coherent states and a regularizing parameter $\ell$. 
As related to the width of the Gaussian, and as explained in \cite{gazeau18}, this parameter has a deep statistical meaning, since it encodes the degree of our confidence in the validity of the classical model.

As mentioned above, one of the interesting outcomes of our approach is the appearance of a smooth PDM on the quantum and semi-classical levels. The effect of PDM in quantum mechanics has been investigated from 
not only a theoretical concern but also a phenomenological requirement. The mostly recent references listed in
\cite{leblond95,quesne,oscar12,mustafa15,evaldo16,bravo16,carinena17,bcosta18} 
give a good idea of the past and current activity in this field. The historical background of the studies of PDM is well-summarized in the respective 
introductions of, for example, \cite{quesne,evaldo16}.
As a matter of fact, a PDM appears naturally when we require the shadow Galilean invariance \cite{leblond95}.
As a theoretical problem associated with PDM, 
we observe that PDM's in classical or quantum mechanics are mostly introduced in a phenomenological way as  functions describing the environment of the considered system. However, their precise forms are not clear and their possible modifications induced by quantization have not been seriously considered so far.
Moreover, in attempting to establish a quantum model from a classical Hamiltonian with PDM, 
one meets the so-called ordering problem of operators in the quantization of the kinetic term where the mass 
becomes a multiplication operator and is not commutable with the momentum operator. 
This is well-known but still an open question in the quantization procedure. 
See, for example, \cite{leblond95}.

This paper is organized as follows.
In Section \ref{CSQ}, we briefly describe the general setting of our quantization procedure. 
In Section \ref{interval}, we implement this procedure in the case of the interval and obtain 
the modified position and  momentum operators, and their respective semi-classical portraits.  The related commutator and  uncertainty relation are analyzed in Section \ref{comrule} by considering the effect of the geometric constraint. 
In Section \ref{sec:2b}, we revisit the classical dynamics with  PDM, first in the continuous case, and next in the discontinuous case 
which appears as a consistent alternative to infinite confinement potentials.
The corresponding quantization is implemented in Section \ref{sec:2c}, 
and the resulting PDM operator and potentials are described in full details. 
Semi-classical phase space portraits of the above operators are examined in Section \ref{semclass}. This leads to interesting observations on the resulting smooth Hamiltonian mechanics and its classical limit.
We conclude in Section \ref{conclu} by listing some interesting problems and generalizations.

\section{Coherent state quantization with geometric restrictions}
\label{CSQ}

Let us consider the quantization of a one-dimensional classical system where the position and the corresponding canonical momentum are denoted by $q$ and $p$, respectively. 
The standard CS quantization is the simplest one among a wide class of integral quantizations. It pertains to  the so-called \textit{elliptic regular quantization}  
\cite{bergaz14,becugaro17}, see also Chapter 11 in \cite{aagbook14}. The operator corresponding to a c-number $f(q,p)$ and acting on the Hilbert space $\mathcal{H}$ of quantum states is accordingly defined by the linear map,
\begin{eqnarray}
f(q,p)\mapsto \hat{A}_{f(q,p)} = \int_{\R^2} \frac{\ud q \ud p}{2\pi  \hbar } \,f(q,p) | q,p \rangle \langle q, p |\, ,
\end{eqnarray}
where kets $| q,p \rangle$ are the standard normalized coherent states in $\mathcal{H}$.
The unit function $f(q,p)= 1$ is mapped to the identity operator $I$ on $\mathcal{H}$, i.e., 
$1 \mapsto  \int_{\R^2} \frac{\ud q \ud p}{2\pi \hbar} \,  | q,p \rangle \langle q, p | = I $.
This is a fundamental property in the CS integral quantization.

The procedure is completed by the so-called semi-classical phase space portrait of $\hat{A}_{f(q,p)}$ defined as its mean value in the same CS, 
\begin{equation}
\label{sclportst}
\check{f}(q,p):= \lg q,p| \hat{A}_{f(q,p)}|q,p\rg= \int_{\R^2} \frac{\ud q^{\prime} \ud p^{\prime}}{2\pi \hbar} \,f(q^{\prime},p^{\prime}) \vert \langle q^{\prime},p^{\prime} | q, p  \rangle\vert^2\, .
\end{equation} 
The regularizing map $f \mapsto \check{f}$ is interpreted as a local averaging of $f(q,p)$ with the probability distribution $(q^\prime,p^\prime) \mapsto \check{\rho}(q^\prime,p^\prime;q,p)$ on the phase space $\R^2$ equipped with the measure $\ud q\ud p/2\pi$.
The CS kets $| q,p \rangle$ read in position representation, for which $\mathcal{H}= L^2(\R,\ud x)$, as
\begin{eqnarray}
\langle x  | q,p \rangle = \left( \frac{1}{\pi \ell^2} \right)^{1/4} e^{-\frac{\ii}{2\hbar} pq} 
e^{\frac{\ii}{\hbar}px} e^{-(x-q)^2/(2\ell^2)}\, ,
\end{eqnarray}
where $\ell$ is a constant which has the dimension of the position. See \cite{bergayou13} for a thorough discussion about the physical significance of this length parameter. The overlap between two CS is given by 
\begin{equation}
\label{overlapCS}
\lg q^{\prime},p^{\prime}|q,p\rg = e^{\ii \frac{pq^{\prime}-qp^{\prime}}{2\hbar}}\, e^{-\frac{(q-q^{\prime})^2}{4\ell^2}}\,  e^{-\ell^2\frac{(p-p^{\prime})^2}{4\hbar^2}}\,,\quad \vert \lg q^{\prime},p^{\prime}|q,p\rg\vert^2=\check{\rho}(q^\prime,p^\prime;q,p)\, . 
\end{equation}

We now consider  classical motions which are geometrically
 restricted to hold in some subset $E_{\mathrm{ps}}$ of the phase space $\R^2$.
Although there is no established procedure, 
we consistently modify the above integral quantization by first truncating all classical observables to $E_{\mathrm{ps}}$ in the following way,
\begin{equation}
\label{trunc}
f(q,p)\mapsto \chi_{E_{\mathrm{ps}}}(q,p)f(q,p)\equiv f_{\chi}(q,p)\, ,
\end{equation}
where $\chi_{E_{\mathrm{ps}}}$ is the characteristic (or indicator) function of set $E_{\mathrm{ps}}$, 
\begin{eqnarray}
\chi_{E_{\mathrm{ps}}} = 
\left\{
\begin{array}{cl}
1 & (q,p)\in E_{\mathrm{ps}} \\
0 & {\rm otherwise}
\end{array} \, .
\right.
\end{eqnarray}
We further proceed with the CS quantization of this not necessarily smooth observable, and obtain the $E$-modified operator,
\begin{eqnarray} 
\label{quantchif}
f_{\chi}(q,p)\mapsto\hat{A}_{f_{\chi}(q,p)} = \int_{\R^2} \frac{\ud q \ud p}{2\pi \hbar} \,\chi_{E_{\mathrm{ps}}}(q,p)\,f(q,p) | q,p \rangle \langle q, p |\, ,
\end{eqnarray}
In the present formulation with geometric constraint, the quantization of $f_\chi(q,p)= \chi_{E_{\mathrm{ps}}}(q,p)$ corresponds to  the ``window'' operator,
\begin{equation}
\label{restId}
 \hat{A}_{\chi_{E_{\mathrm{ps}}}(q,p)} = \int_{E_{\mathrm{ps}}} \frac{\ud q \ud p}{2\pi \hbar} \, | q,p \rangle \langle q, p |\equiv \hat{B}_{E_{\mathrm{ps}}}\, .
\end{equation}
Note that the  Hilbert space  in which act these ``$E$-modified'' operators is left unchanged. Thus, in position representation, one continues to deal with  $\mathcal{H}= L^2(\R, \ud x)$.
Nevertheless, our approach gives rise to a smoothing of the constraint boundary, 
i.e., a ``fuzzy'' boundary, and also a smoothing of all discontinuous restricted observable $f_\chi(q,p)$ introduced in the quantization map \eqref{quantchif}. 
Indeed, there is no mechanics outside the strip defined by the position interval constraint  on the classical level. 
It is however not the same on the quantum level since our quantization method allows to go beyond the boundary in a rapidly decreasing smooth way.

Consistently,  the semi-classical phase space portrait of the operator \eqref{quantchif} 
\begin{equation}
\label{semclassA}
 \check{A}_{f_\chi(q,p)}(q,p)= \int_{E_{\mathrm{ps}}} \frac{\ud q^{\prime} \ud p^{\prime}}{2\pi \hbar} \,f(q^{\prime},p^{\prime}) \vert \langle q^{\prime},p^{\prime} | q, p  \rangle\vert^2\, , 
\end{equation}
should be found to be concentrated on the classical $E_{\mathrm{ps}}\subset \R^2$. Such a function should be viewed as a new classical observable defined on the full phase space $\R^2$ where  $q$ and $p$ keep their status of canonical variables. 

Thus, we have here the interesting sequence\\
 \begin{equation}
\label{regseq}
\mathrm{virtual} \ f(q,p) \rightarrow \  \mathrm{truncated} \ f_\chi(q,p) \rightarrow\  \mathrm{quantum}\  \hat A_{f_{\chi}}
\rightarrow\ \mathrm{regularised} \ \check f_\chi(q,p)\,,
\end{equation}
allowing to establish a semi-classical dynamics \textit{\`a la} Klauder \cite{klauder12},  mainly concentrated on $E_{\mathrm{ps}}$. 

Note that the truncated position and momentum $(q_\chi,p_\chi)$ still behave as canonical variables in the subset $E_{\mathrm{ps}}$.
The semi-classical portraits of their respective quantum counterparts are defined even outside the subset $E_{\mathrm{ps}}$
and thus the canonical properties of the operators $(\hat{A}_{q_\chi}, \hat{A}_{p_\chi})$ should be investigated carefully as is done in the following two sections.

\section{Position and momentum operators in an interval and their semi-classical portraits}
\label{interval}

Because we are interested in constraints in the configuration space only,  we examine  geometric restrictions $E_{\mathrm{ps}}=E\times \R$ where $E\subset \R$ in the configuration space and we put to simplify notations $\chi_{E_{\mathrm{ps}}}(q,p) = \chi_E\otimes 1(q,p)\equiv \chi_E(q)$. Actually, we restrict the study to the bounded open interval $E= (a,b)$, i.e., 
\begin{eqnarray}
\chi_E(q)= \chi_{(a,b)}(q) =\Theta(q-a) - \Theta(q-b)\, ,
\end{eqnarray}
where $\Theta(q)=  \chi_{(0,+\infty)}(q)$ is the Heaviside function. Hence, we have  at our disposal four parameters for the bounded interval case. Two of them, $a$ and $b$, are associated with classical geometrical constraints, one, $\ell$, is introduced through the quantization procedure, 
and the last one, $\hbar$, is for the quantum model. Lengths $a$, $b$, and $\ell$ will be expressed in terms of a certain unit $q_0$, and  we consider in our study fixed values of $b/q_0=a/q_0+10$ while $a/q_0$ takes three different values, $a=0,2$ and $4$. 
Occasionally, we also consider  the semi-bounded $E= (0,\infty)$, i.e., the positive half-line 
by setting $a=0$ and $b=\infty$.  However our study is mainly centered on the bounded case.

A crucial  aspect of any quantum model is its classical limit. In the present case, the latter is carried out through the simultaneous limits \cite{bergaz14}
\begin{equation}
\label{classlim}
\hbar \to 0\,, \quad \ell \to 0\, , \quad \frac{\hbar}{\ell} \to 0\, .
\end{equation}

At this point, we should be aware that the motion in our bounded or semi-bounded geometry is not identified 
with the bounded motion induced by a confinement potential such as two infinite potential walls, although there are similar features on the classical level as is discussed in Section \ref{sec:2b}.

\begin{figure}[t]
\begin{center}
\includegraphics[scale=0.3]{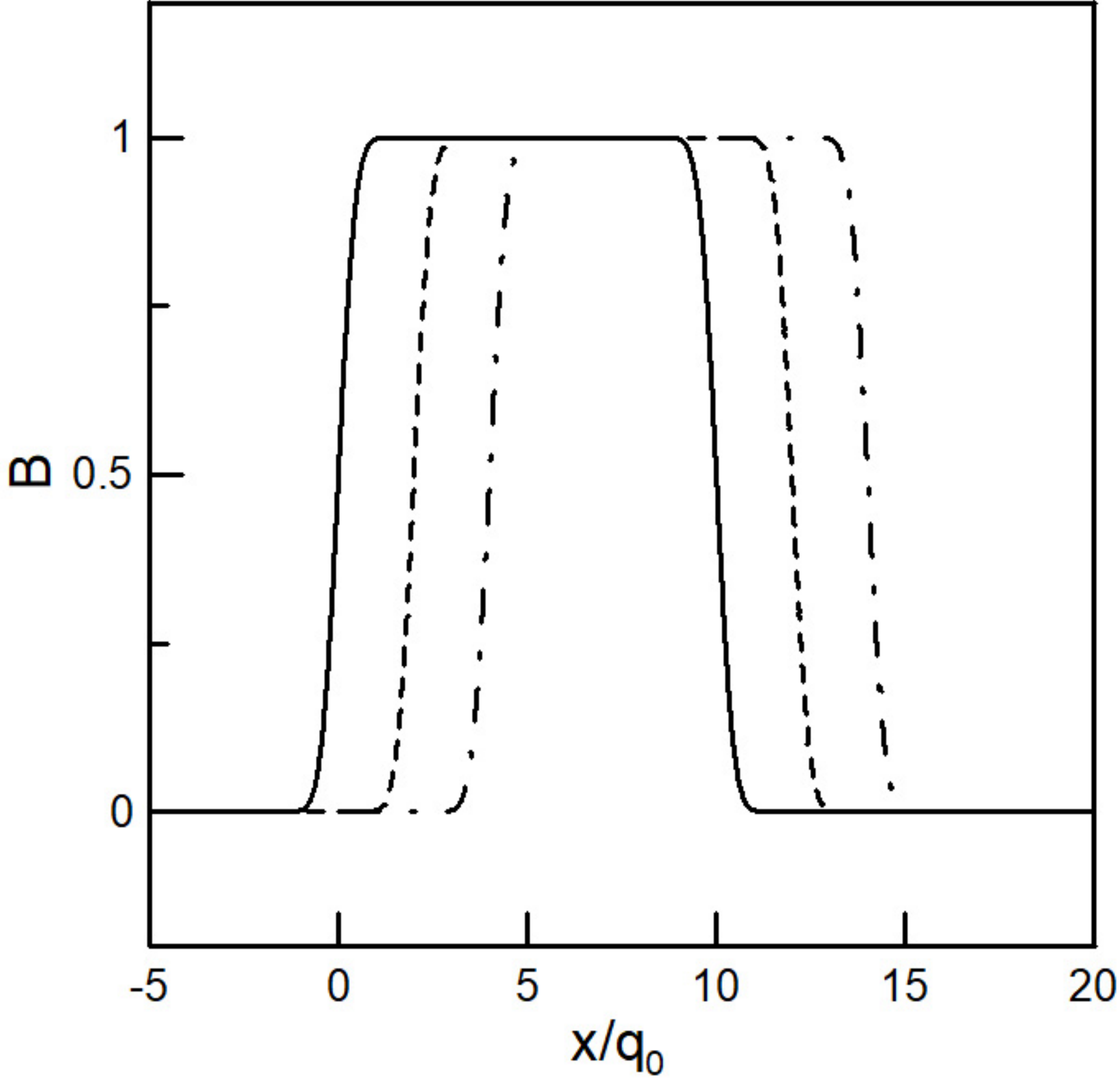}
\includegraphics[scale=0.3]{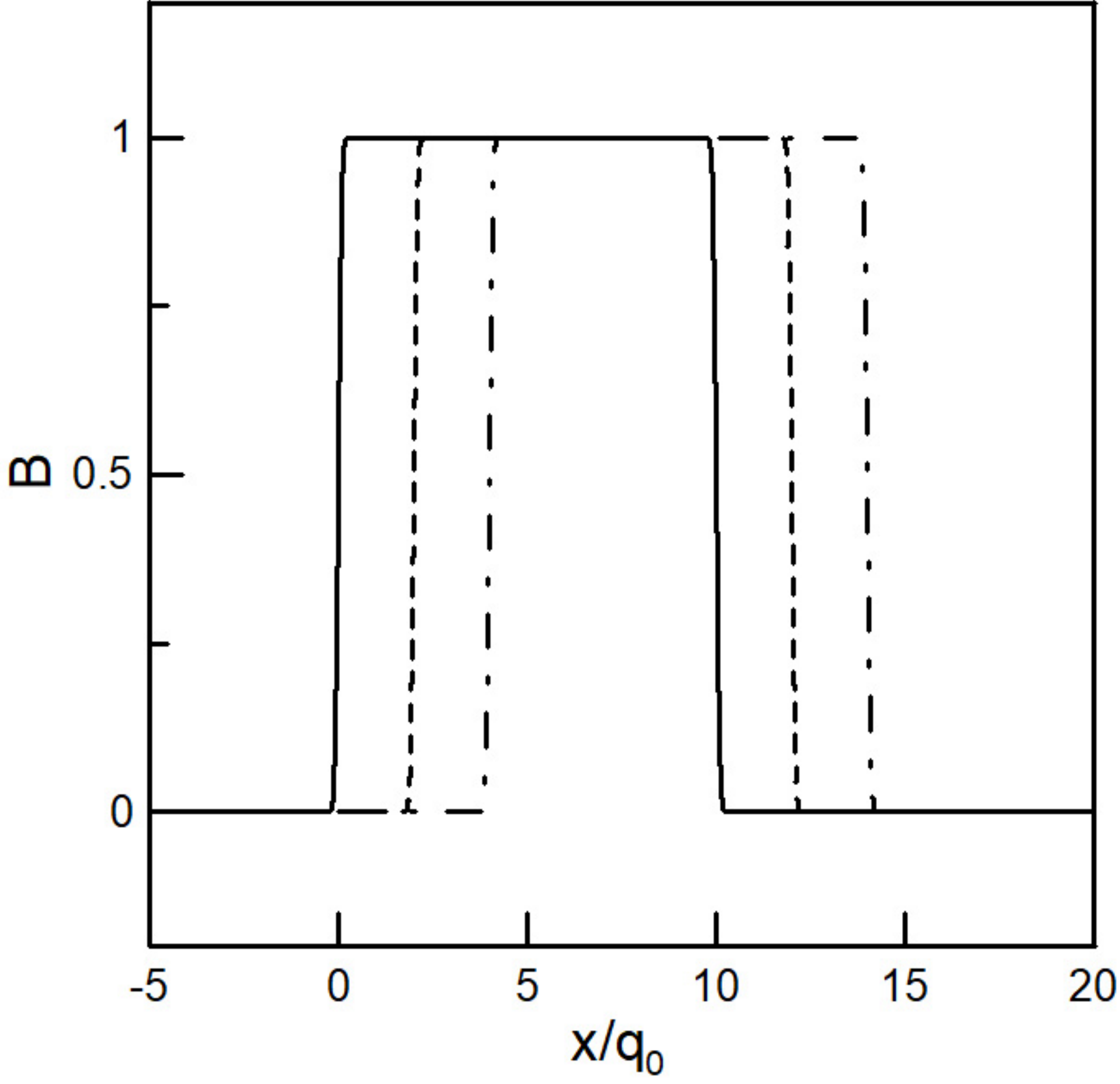}
\end{center}
\caption{The behaviors of the function $B_{\ell}(x,a,a+10 q_0)$ for different values of $a$.  
The quantization length $\ell$ is chosen to be $\ell = 0.5q_0$ (left panel) and $\ell = 0.1q_0$ (right panel).  
The solid, dashed and dot-dashed lines represent the results of $a/q_0=0,2$ and $4$, respectively.}
\label{fig:B}
\end{figure}

The semi-classical phase space portrait of the window operator (the characteristic function) is given by
\begin{equation}
\label{fct1qab}
{\check{B}}_{(a,b)}(q,p)= \frac{1}{2\pi \hbar} \int_a^b \ud q^{\prime} \int_{-\infty}^{+\infty}\ud p^{\prime}\, \vert \langle q^{\prime},p^{\prime} | q, p  \rangle\vert^2 
= B_{\ell}\left(\frac{q}{\sqrt{2}},\frac{a}{\sqrt{2}},\frac{b}{\sqrt{2}}\right)= B_{\sqrt{2}\ell}(q,a,b) \, , 
\end{equation}
where
\begin{eqnarray}
B_{\ell}(x,a,b) = \frac{1}{2}\left[\erfc \left( \mathnormal{\frac{ a-x }{\ell}} \right) 
- \erfc \left( \mathnormal{\frac{ b-x }{\ell}} \right)\right]\, ,
\end{eqnarray}
is expressed in terms of the complementary error function $\erfc$ \cite{abraste72}
\begin{eqnarray}
\erfc (x) 
= 
\frac{2}{\sqrt{\pi}} \int^\infty_x \ud t\, e^{-t^2}\, .
\end{eqnarray}  
This function is smooth, symmetric with respect to the middle of the interval $[a,b]$,
\begin{equation}
\label{ }
B_{\ell}(x,a,b) = B_{\ell}(a+b -x,a,b)\, . 
\end{equation}
and vanishes rapidly outside the interval, 
i.e., it belongs to the Schwartz space $\mathcal{S}$ of smooth rapidly decreasing functions on the line. 
Note the value at the endpoints:
\begin{equation}
\label{ }
B_{\ell}(a,a,b)= B_{\ell}(b,a,b)= \frac{1}{2} - \frac{1}{2}\erfc \left( \frac{a-b}{\ell}\right)\,.  
\end{equation}

One can easily confirm that this semi-classical portrait of the window operator represents the smooth regularization of the 
characteristic function. In fact, in the vanishing limit of $\ell$, we observe 
\begin{equation}
\label{regstepab}
B_{\ell} (x,a,b)  
\underset{\ell \to 0}{\to} \Theta(x-a) -\Theta(x-b)= \cab(x)\,.
\end{equation}
Consistently, the  derivative of $B_{\ell} (x,a,b)$ is the well-known Gaussian regularization of the Dirac distribution:
\begin{equation}
\label{regdirab}
\frac{\ud}{\ud x}B_{\ell} (x,a,b) =  \frac{1}{\sqrt{\pi}\ell}\left(e^{-\frac{(x-a)^2}{\ell^2}}-e^{-\frac{(x-b)^2}{\ell^2}}\right)
\underset{\ell \to 0}{\to} \delta(x-a) -\delta(x-b)\equiv \delta_a(x)- \delta_b(x) \,.
\end{equation}

The above results are directly applicable to a semi-bounded geometry, like the positive half-line, finding
\begin{eqnarray}
\label{chBi}
{\check{B}}_{(0,\infty)} (q,p) = B_{\sqrt{2\ell}} (x,0,\infty) 
\, ,
\end{eqnarray}
with the difference that this smooth function is not rapidly decreasing on the right. Indeed, 
one checks  that $\lim_{x \rightarrow +\infty} B_{\ell} (x,0,\infty) = 1$ while $\lim_{x \rightarrow -\infty} B_{\ell} (x,0,\infty) = 0$. 
In the vanishing limit of $\ell$, again we find 
\begin{equation}
\label{regstep}
B_{\ell} (x,0,\infty)  
\underset{\ell \to 0}{\to} \Theta(x)\, ,
\end{equation}
and the  derivative of $B_{\ell} (x,0,\infty)$ is expressed as 
\begin{equation}
\label{regdir}
\frac{\ud}{\ud x}B_{\ell} (x,0,\infty)= \frac{1}{\sqrt{\pi}\ell}e^{-\frac{x^2}{\ell^2}} \underset{\ell \to 0}{\to} \delta_{0}(x)\,.
\end{equation}

As was already defined, the quantization of $f(q,p)=1$ leads to the bounded self-adjoint window operator which is multiplicative, 
\begin{eqnarray}
\hat{A}_{\chi_{(a,b)}}  \phi(x) = B_{\ell}(x,a,b)  \phi(x)\, , 
\end{eqnarray}
where $\phi(x)$ is an arbitrary wave function in $L^2(\R,\ud x)$.
The behaviors of $B_{\ell}(x,a,b)$ are shown in Fig.\  \ref{fig:B}. 
The left and right panels show the results with $\ell = 0.5q_0$ and $\ell = 0.1q_0$, respectively.
The solid, dashed and dot-dashed lines represent the results of $a/q_0=0,2$ and $4$, respectively.
One can observe the rapid transition from $0$ to $1$ (from $1$ to $0$) at the endpoints $a$ ($b$). 
The classical discontinuities included in $\chi_E (q,p)$ are smoothed because of the Gaussian nature of the coherent states employed in the quantization procedure. 
In fact, the rapid transitions are enhanced as $\ell$ decreases. 
That is, $B_{\ell}(x,a,b)$ goes to $\chi_{(a,b)}$ in the classical limit $\ell \to 0$, as expected. 
In the case of the positive half-line, one has just to keep the behavior at $a=0$.

\begin{figure}[h]
\includegraphics[scale=0.3]{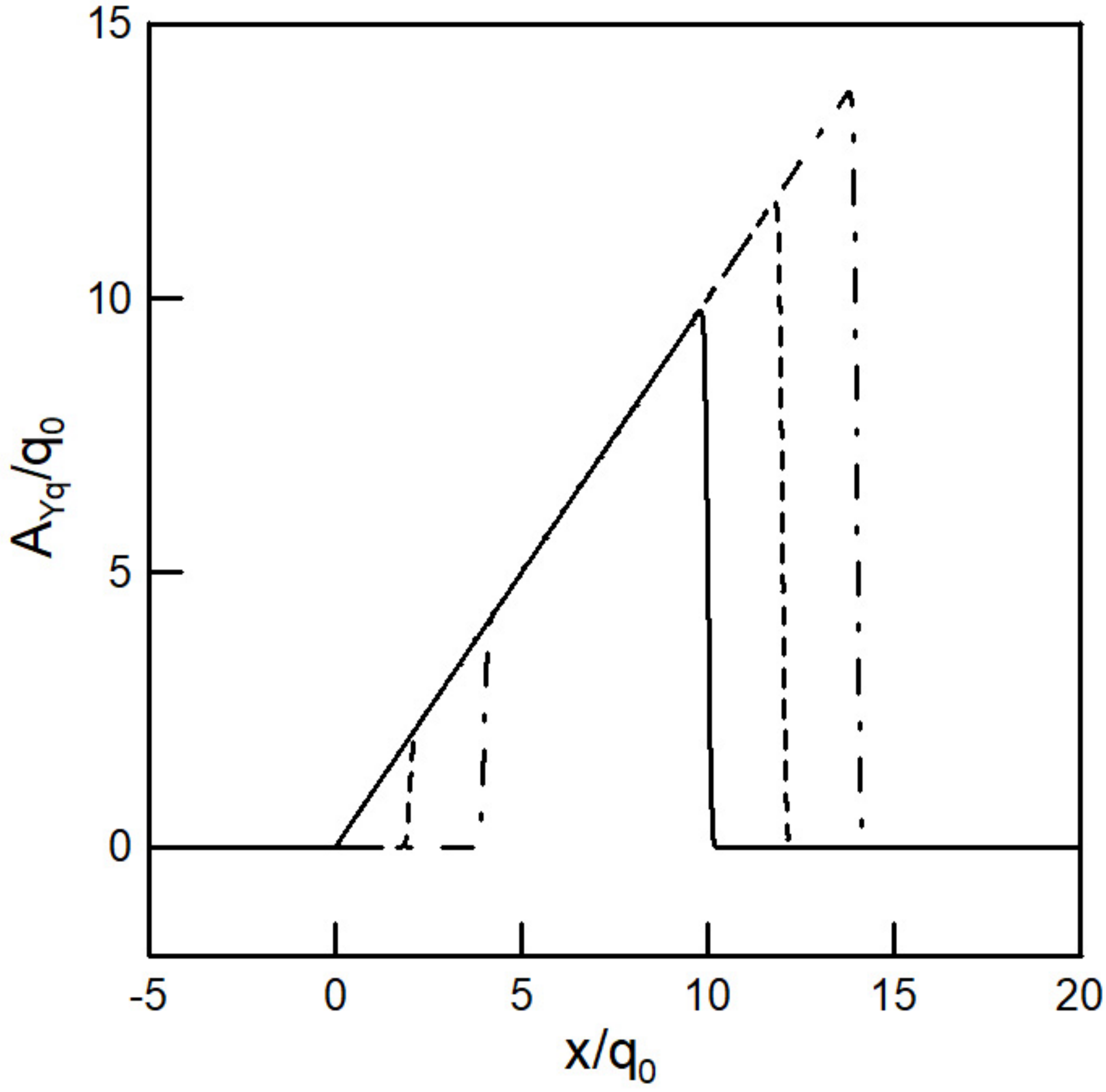}
\caption{The spectral behaviors of the $E$-modified position operator $\hat{A}_{q_\chi}$.
We choose $\ell = 0.1 q_0$. The solid, dashed and dot-dashed lines represent the results of $a/q_0=0,2$ and $4$, respectively. } 
\label{fig:Ayq1}
\end{figure}

In a similar fashion, the $E$-modified  position and momentum operators, respectively, are given by  
\begin{eqnarray}
\hat{A}_{q_\chi} \phi(x) &=& \left( x B_{\ell}(x,a,b) + \ell^2 \frac{B_{\ell}^{(1)} (x,a,b)}{2} \right) \phi(x) \nonumber \\
&\equiv& Q(x)\phi(x)\,, \\
\hat{A}_{ p_\chi} \phi(x) &=& \frac{1}{2} \left\{ B_{\ell}(x,a,b),  -\ii\hbar  \frac{\ud}{\ud x} \right\} \phi(x) \nonumber \\
\label{restmom}&=& -\ii \hbar\left(B_{\ell}(x,a,b) \frac{\ud}{\ud x} + \frac{1}{2} B_{\ell}^{(1)} (x,a,b)\right) \phi(x)\, ,
\end{eqnarray}
where $\{\ \}$ is the anti-commutator and 
\begin{eqnarray}
B_{\ell}^{(n)} (x,a,b) 
&=& \frac{\ud^n}{\ud x^n} B_{\ell}(x,a,b)\,.
\end{eqnarray}
Therefore, we have the relation between these operators and the usual position and momentum operators $\hat q\phi(x) = x\phi(x)$,  and $\hat p\phi(x)= -\ii \hbar \dfrac{\ud}{\ud x} \phi(x)$:
\begin{equation}
\label{qpAqAp}
\hat{A}_{q_\chi} = \hq B_{\ell}(\hq,a,b) + \ell^2 \frac{B_{\ell}^{(1)} (\hq,a,b)}{2}=F(\hat q)\, , 
\quad \hat{A}_{p_\chi} 
= \frac{1}{2} \left\{ B_{\ell}(\hq,a,b),  \hp \right\}\, .  
\end{equation} 
We notice that the $E-$modified position and momentum operators, $\hat{A}_{q_\chi}$ and $\hat{A}_{p_\chi} $, 
reduce to the standard ones $ \hq$ and $ \hp$, inside the interval $(a,b)$, more precisely, there where  
$B \approx 1$. 

The $E$-modified position operator $\hat{A}_{q_\chi}$ is bounded self-adjoint, and its spectral measure is 
\begin{equation}
\label{specmeas}
\ud Q(x)= \left( B_{\ell}(x,a,b) + x B^{(1)}_{\ell}(x,a,b) + \ell^2 \frac{B_{\ell}^{(2)} (x,a,b)}{2}\right) \ud x\,. 
\end{equation}
The spectral behaviors of the $E$-modified position operator $\hat{A}_{\cab q}$ are shown in Fig.\ \ref{fig:Ayq1}.
Choosing $\ell = 0.1 q_0$, the solid, dashed and dot-dashed lines represent 
the results of $a/q_0=0,2$ and $4$, respectively.
One can easily see that the spectrum of $\hat{A}_{\cab q}$ behaves as a linear function of $x$ deeply 
inside the interval $[a,b]$ while it becomes fast negligible outside the interval 
as is shown in Fig.\ \ref{fig:Ayq1}. 
This means that the spectral measure of $\hat{A}_{\cab q}$ is essentially concentrated on the interval $(a,b)$, and that
\begin{equation}
\label{specconc1}
  \int^{d}_{c} \ud Q \approx d-c\, , 
\end{equation}
for $a<c\leq d<b$, and that 
\begin{equation}
\label{specconc2}
  \int_{E} \ud Q \approx 0\, , 
\end{equation}
for any set $E$ such that $E\cap(a,b)=\emptyset$ or $E\cap(a,b)=(a,b)$.
We check  that the scale of the smooth approximation to the discontinuity near the boundaries is characterized by $\ell$ as expected from Eq.\;\eqref{regstep}.

Concerning the $E$-modified momentum operator \eqref{restmom}, it is symmetric by construction and  
unbounded (it is approximately $\hat p$ for $x$ deeply in the interval $(a,b)$). 
As a symmetrized product with the multiplication operator $B_{\ell}(\hat x,a,b)\in \mathcal{S}$,  it is, like $\hat p$,  
essential self-adjoint in $L^2(\R,\ud x)$ (both have same core domain \cite{reedsimon75}). 
At this point, one should remind that the momentum operator for the quantum motion in the interval $(a,b)$ 
is not essentially self-adjoint, but has a continuous set of self-adjoint extensions, due to the infinite 
choices of boundary conditions in defining its domain. In the present case, our approach allows 
to get around 
the {ambiguity of the boundary conditions since the discontinuous characteristic function of the interval 
is replaced by the smooth function $B_{\ell}(x,a,b) $ which rapidly vanishes outside the interval, 
and since the Hilbert space 
of wave functions is $L^2(\R,\ud x)$ and not $L^2((a,b),\ud x)$.

\begin{figure}[h]
\begin{center}
\includegraphics[scale=0.3]{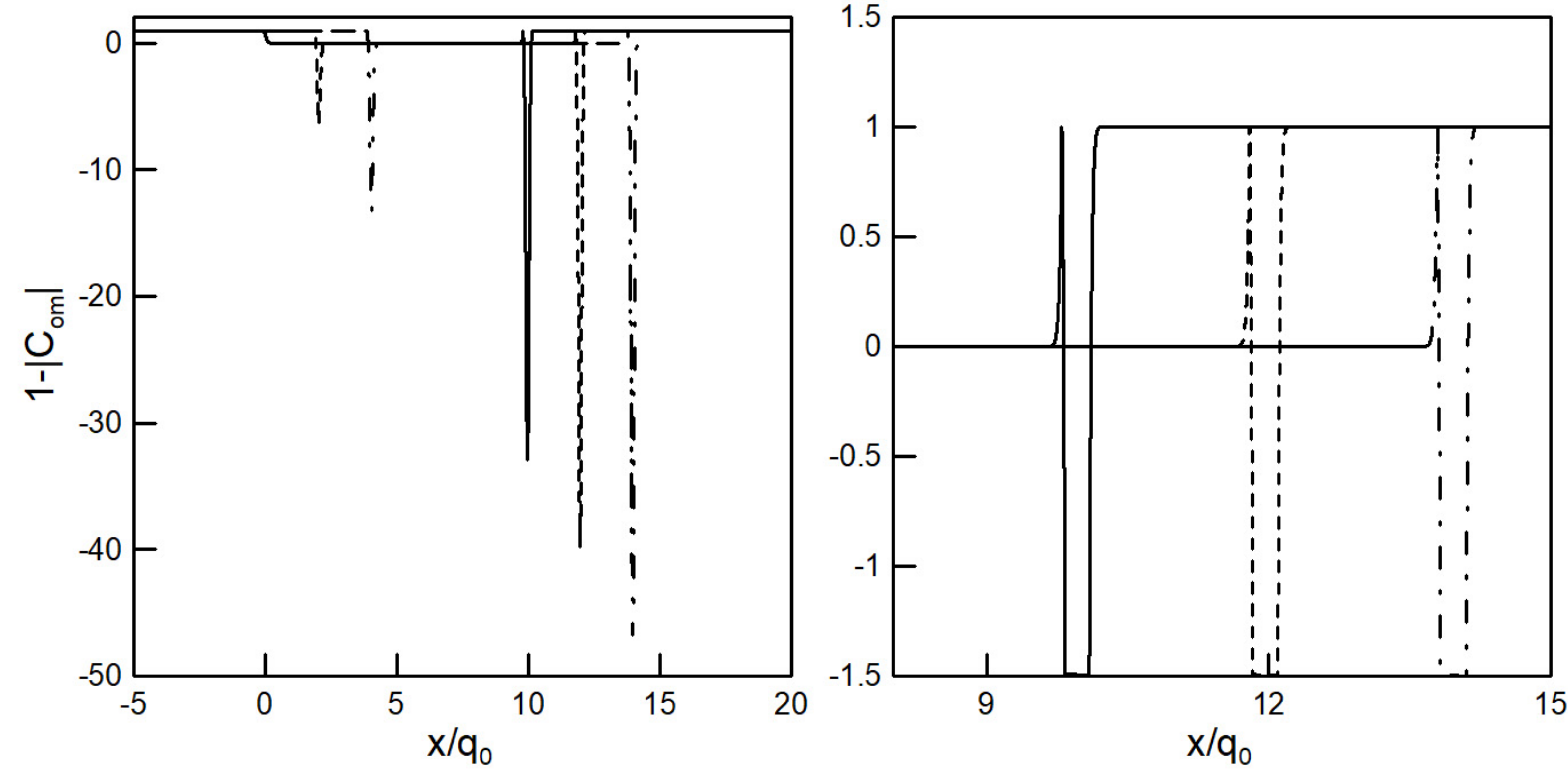}
\end{center}
\caption{The departure from $1$ of the absolute value of the commutator $ |{\rm C}_{\rm om}(x,a,a+10 \ell)|$ with $\ell = 0.1 q_0$ is shown on the left panel. The solid, dashed and dot-dashed lines represent the results of $a/q_0=0,2$ and $4$, respectively. 
The figure on the right panel is the enlarged figure of the left one for the domain of $8 \le x/q_0 \le 15$.
} 
\label{fig:Ayq2}
\end{figure}
We end this section by giving the expressions of the semi-classical phase space portraits of the $E$-modified position and momentum operators using Eq.\ (\ref{sclportst}). They are respectively given by the smooth functions 
\begin{align}
\label{semclq}
\check{q}_{\chi} (q,p)  &= q B_{\sqrt{2}\ell} (q,a,b) + \ell^2 B^{(1)}_{\sqrt{2}\ell} (q,a,b)\, ,   \\
 \label{semclp}  \check{p}_{\chi} (q,p) & =  p B_{\sqrt{2}\ell}(q,a,b)\, ,
\end{align}
which are rapidly decreasing out of the strip $(a,b)\times \R$ at fixed $p$, and whose limits at $\ell =0$ are respectively $q_\chi$ and $ p_\chi$, as expected.

\section{The $E$-modified canonical rule}
\label{comrule}

As was mentioned,  $(q_\chi,p_\chi)$ are still canonical within the interval. Therefore the $E$-modified position and momentum operators behave as 
canonical variables inside the bounded geometry, while we observe deviation outside of the constraint. Indeed, 
the commutator of the $E$-modified position and momentum operators read 
\begin{eqnarray}
[\hat{A}_{q_\chi}, \hat{A}_{p_\chi}] = \ii \hbar {\rm C}_{\rm om}(\hq,a,b)\,, \label{eqn:commu}
\end{eqnarray}
where
\begin{eqnarray}
\label{Comx}
{\rm C}_{\rm om} (x,a,b) 
&=&  B_{\ell}(x,a,b) \left( B_{\ell}(x,a,b) + xB_{\ell}^{(1)}(x,a,b) + \frac{\ell^2}{2} B_{\ell}^{(2)}(x,a,b) \right) \nonumber \\
&=& B_{\ell}(x,a,b) \frac{dQ}{dx}(x)
\, .
\end{eqnarray}
In Fig.\ \ref{fig:Ayq2} is shown the departure 
from $1$ of the absolute value of ${\rm C}_{\rm om}(x,a,b)$ choosing $\ell = 0.1 q_0$. The solid, dashed and dot-dashed lines represent the results of $a/q_0=0,2$ and $4$, respectively.
If the $E$-modified position and momentum operators were canonical, the right-hand side of the commutator (\ref{eqn:commu}) should be $i\hbar$, i.e., ${\rm C}_{om} =1$. In fact, it is satisfied deeply inside the interval $(a,b)$. Near and outside the endpoints of the interval, however, we observe oscillations associated with non-canonical behaviors, which are enhanced as the positions of endpoints $a$ and $b$ increase.

In the vanishing limit of $\ell$, we find 
\begin{equation}
\label{eqn:comsemi1}
{\rm C}_{\rm om} (q,a,b)
\underset{\ell \to 0}{\to} \cab(q)+``\cab(q) \left( a\delta_a(q)- b\delta_b(q)\right)"\, , 
\end{equation}
where we used $(\cab(q))^2 = \cab(q)$.
Here appears  the ill-defined product $``\delta_{a(b)}(x)\Theta(x)"$. 
For instance, we could try to use the regularization \eqref{regstep} of $\Theta(x)$ to consider the following action  on a test function $\varphi(x)$, 
\begin{equation}
\label{Hdelta}
\begin{split}
\int_{-\infty}^{+\infty}``\delta(x)\Theta(x)" \varphi(x) \ud x&:= \lim_{_{\ell \to 0}} \int_{-\infty}^{+\infty}\delta(x) B_{\ell} (x,0,\infty)\varphi(x) \ud x\\ & =B_{\ell} (0,0,\infty)\varphi(0) 
= \frac{1}{2}\varphi(0)\,, \ \mbox{i.e.}\  ``\delta(x)\Theta(x)":= \frac{1}{2} \delta(x)\,.
\end{split}
\end{equation}
Therefore, if we accept this definition, Eq.\ (\ref{eqn:comsemi1}) becomes
\begin{eqnarray} 
{\rm C}_{\rm om} (q,a,b)
\underset{\ell \to 0}{\to} \cab(q)+\frac{1}{2} \left( a\delta_a(q)- b\delta_b(q)\right)\,. 
\end{eqnarray}
Again we observe that the canonical property of the $E$-modified position and momentum operators holds only inside the interval.

To see the consistency of the above result, let us calculate the Poisson bracket for the semi-classical quantities, 
$\check{q}_{\chi}$ and $\check{p}_{\chi}$ which are  smooth observables on $\R^2$.
Note that these variables are not canonical variables. 
We have
\begin{eqnarray}
\label{PBscl}
\left\{\check{q}_{\chi},\check{p}_{\chi} \right\}_{PB}
&=&
\frac{\partial \check{q}_{\chi}}{\partial q} \frac{\partial\check{p}_{\chi}}{\partial p} 
- \frac{\partial \check{q}_{\chi}}{\partial p} \frac{\partial \check{p}_{\chi}}{\partial q}   \nonumber \\
&=& 
B_{\sqrt{2}\ell}(q,a,b) \left( B_{\sqrt{2}\ell}(q,a,b) + qB_{\sqrt{2}\ell}^{(1)}(q,a,b)
+ \ell^2 B^{(2)}_{\sqrt{2}\ell} (q,a,b)
\right)\, ,
\end{eqnarray}
and one can easily see that the vanishing limit of $\ell$ of the right-hand side of this equation reproduces the classical limit of ${\rm C}_{om} (q,a,b)$ given in Eq.\ \eqref{Comx}. 
Therefore, using the correspondence $\frac{1}{i\hbar} [\, , \, ] \mapsto\{\, , \, \}  $, we can see in the classical limit, 
\begin{eqnarray}
 \frac{1}{i\hbar}  \lim_{\ell \rightarrow 0} [\hat{A}_{q_\chi}, \hat{A}_{p_\chi}]\mapsto  \lim_{\ell \rightarrow 0} \left\{\check{q}_{\chi},\check{p}_{\chi} \right\}_{PB}
\, .
\end{eqnarray}

\begin{figure}[h]
\begin{center}
\includegraphics[scale=0.3]{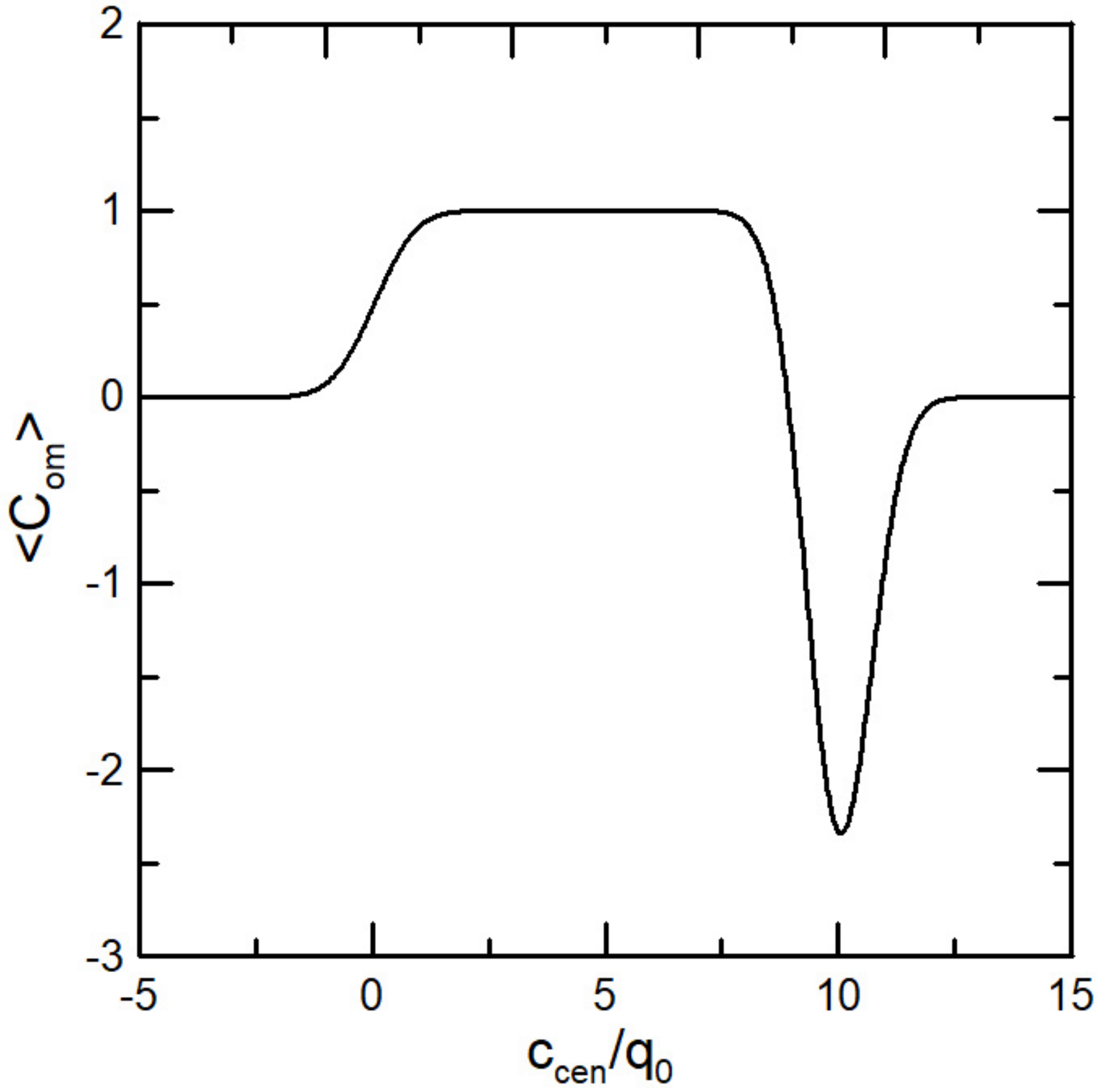}
\end{center}
\caption{The behavior of $\langle C_{om} \rangle$ is shown as a function of $c_{cen}$.
We set $\ell = 0.1 q_0$, $a=0$ and $b = 10 q_0$.
}
\label{fig:unc-int}
\end{figure}

To better understand the physics attributed to the modification of the canonical commutator Eq.\ (\ref{eqn:commu}), 
the uncertainty relation between the position and momentum is calculated as
\begin{eqnarray}
\label{uncert}
(\Delta \hat A_{q_\chi}) (\Delta {\hat A}_{p_\chi}) \ge \frac{\hbar}{2} |{\lg {\rm C}_{\rm om}(\hq,a,b)\rg}|\, ,
\end{eqnarray}
where we introduced the standard deviation which is defined for an arbitrary operator $\hat{f}$ as 
\begin{eqnarray}
\Delta f = \sqrt{\langle \hat{f}^2 \rangle - \langle \hat{f} \rangle^2}\, ,
\end{eqnarray}
and expected values $\lg\cdot\rg$ are calculated with an arbitrary quantum state. 
As is predicted from Fig.\ \ref{fig:Ayq2}, when the wave function is located in the interval $[a,b]$, 
$C_{om} (x,a,b) =1$ and thus the standard minimum uncertainty is reproduced.

The modification of the minimum uncertainty can be observed when the wave function stays near the endpoints of boundaries.
To see this, let us consider the following Gaussian wave function to calculate $\langle C_{om} \rangle$, 
\begin{eqnarray}
\psi(x,c_{cen}) =\frac{1}{\sqrt{q_0} \pi^{1/4}} e^{-(x-c_{cen})^2/(2q^2_0)}\, ,
\end{eqnarray}
where $c_{cen}$ is a parameter to characterize the center of the Gaussian distribution.
In Fig.\ \ref{fig:unc-int}, 
the behavior of $\langle C_{om} \rangle$ is shown as a function of $c_{cen}$, 
setting $\ell = 0.1 q_0$.
As was mentioned, the standard minimum uncertainty $\langle C_{om} \rangle =1$ is reproduced 
when the position of the center of the Gaussian wave function stays deeply inside the interval $(0, 10 q_0)$. 
On the other hand, in certain regions near the endpoints, the minimum uncertainty can be deviated away from the standard value, $\hbar/2$.

\section{Classical Hamiltonian }
\label{sec:2b}

Before investigating the quantum dynamics of a free particle with mass $m$ and confined 
in a bounded or semi-bounded geometry,
let us consider its  classical model. 
Geometric constraints are sometimes taken into account by introducing potentials. 
In the present calculation, however, we have considered that the quantities of the phase space in the bounded or semi-bounded geometry are multiplied by  $\chi_{(a,b)}(q)$. Therefore if we have the Hamiltonian $H$ in the unbounded geometry, the corresponding truncated quantity $H_{\chi}$ in the bounded or semi-bounded geometry should be expressed as 
\begin{eqnarray}
\label{hhchi}
H_{\chi }(q,p) = \chi_{(a,b)}(q)\,  H(q,p)\, .
\end{eqnarray}
Usually, the kinetic term of Hamiltonian is anti-proportional to the mass of particle.
Then we notice that the geometrical restriction encoded by the characteristic function $\chi_{(a,b)}(q)$ can be equivalent to imposing the following discontinuous PDM on the classical level to the standard Hamiltonian,
\begin{equation}
\label{posdepm}
m_{1/\chi}(q)
:= \frac{m}{\chi_{(a,b)}(q)} 
=\left\lbrace\begin{array}{cc}
   \infty   & q\notin (a,b)  \\
    m  &   a<q<b
\end{array} \right.\, . 
\end{equation}
This is an interesting alternative to the usual approach to the motion of a particle of constant mass $m$ 
trapped by infinite potential walls. However, we have to be very cautious in implementing  the Lagrangian or Hamiltonian formalism to this case,  due to the singular nature of this function. The natural alternative is to proceed first with the smooth regularization of the model yielded by its semi-classical phase space portrait described in the previous sections, and taking eventually its classical limit. Then we will show that both models are, to some extent, equivalent.  

Hence, let us first consider a classical system described by the following Lagrangian for a particle with an arbitrary smooth PDM $m(q)$,
\begin{eqnarray}
\label{laggen}
L(q,\dot{q}) = \frac{m(q)}{2}\dot{q}^2 - V(q)\, ,
\end{eqnarray}
where $V(q)$ is a potential term.
One can easily confirm that the Euler-Lagrange equation is given by 
\begin{eqnarray}
\label{moteq}
m(q) \ddot{q} + \frac{\ud V}{\ud q}(q) 
+ \frac{1}{2}\frac{\ud m}{\ud q}(q)\dot{q}^2 =0\, ,
\end{eqnarray}
and that the following energy of this system is conserved, 
\begin{eqnarray}
E = \frac{m(q)}{2}\dot{q}^2 + V(q)\, .
\end{eqnarray}

Let us now discuss the canonical formulation. 
The canonical momentum is defined by 
\begin{eqnarray}
\label{pfunctionofq}
p = \frac{\partial L}{\partial \dot{q}} = m(q) \dot{q}.
\end{eqnarray}
and it is straightforward to check that the set of $(q,p)$ form canonical variables, satisfying $\{ q, p \}_{PB} = 1$.
The Hamiltonian is defined by the Legendre transformation, 
\begin{eqnarray}
\label{Hvar}
H(q,p) = \dot{q}p - L = \frac{p^2}{2m(q)} + V(q)\, .
\end{eqnarray}
The canonical equations follow from this expression: 
\begin{eqnarray}
\label{dotq}
\dot{q} 
&=& \{ q, H \}_{PB} 
= \frac{p}{m(q)}, \\
\label{dotp} \dot{p}
&=& \{ p, H \}_{PB} = - \frac{\ud V}{\ud q}(q) 
+ \frac{p^2}{2m^2(q)}
\frac{\ud m}{\ud q}(q)\, .  
\end{eqnarray}
For the sake of later convenience, 
we define the force exerted on the particle as the observable
\begin{equation}
\label{classforce}
F(q,p):= m(q) \ddot{q}  
= - \frac{\ud V}{\ud q}(q) - \frac{\dot{q}^2}{2} \frac{\ud m}{\ud q}(q)
= - \frac{\ud V}{\ud q}(q) - \frac{p^2}{2m^2(q)} \frac{\ud m}{\ud q}(q)\, .
\end{equation}
We note that the extra term due to the PDM is the opposite to the second term in the expression \eqref{dotp}. Thus, the Newton law $F=\dot p$ for constant mass loses its validity in the PDM case.

 In the case of our example of constrained geometry, the Lagrangian \eqref{laggen} becomes
\begin{eqnarray}
\label{laggenchi}
L_\chi(q,\dot{q}) = \frac{m_{1/\chi}(q)}{2}\dot{q_\chi}^2 - V_\chi(q)\, ,
\end{eqnarray}
and, with $p_\chi = \dfrac{\partial L_\chi}{\partial \dot{q_\chi}}=m_{1/\chi}(q)\dot{q_\chi}$,  the Hamiltonian \eqref{Hvar} reads as
\begin{equation}
\label{Hvarchi}
H_\chi(q,p) = \dot{q_{\chi}}p_\chi - L_\chi = \frac{p_\chi^2}{2m_{1/\chi}(q)} + V_\chi(q)\, .
\end{equation}
Now the difficulties might arise from the computation of the derivative of the singular  PDM in the applications of Equations \eqref{dotp} and \eqref{classforce}. A solution to this question will be given in  Section \ref{semclass}.

\section{Quantum Hamiltonian and Schr\"{o}dinger equation} \label{sec:2c}

We now consider a free particle with mass $m$ constrained to move in the interval $(a,b)$. According to our definition \eqref{trunc}, its Hamiltonian is defined as the  truncated expression
\begin{equation}
\label{truncH}
H_{\chi}(q,p) = \chi_{(a,b)} (q) \frac{p^2}{2m} =  \frac{p_\chi^2}{2m_{1/\chi}(q)}\,.
\end{equation}

\begin{figure}[t]
\begin{center}
\includegraphics[scale=0.3]{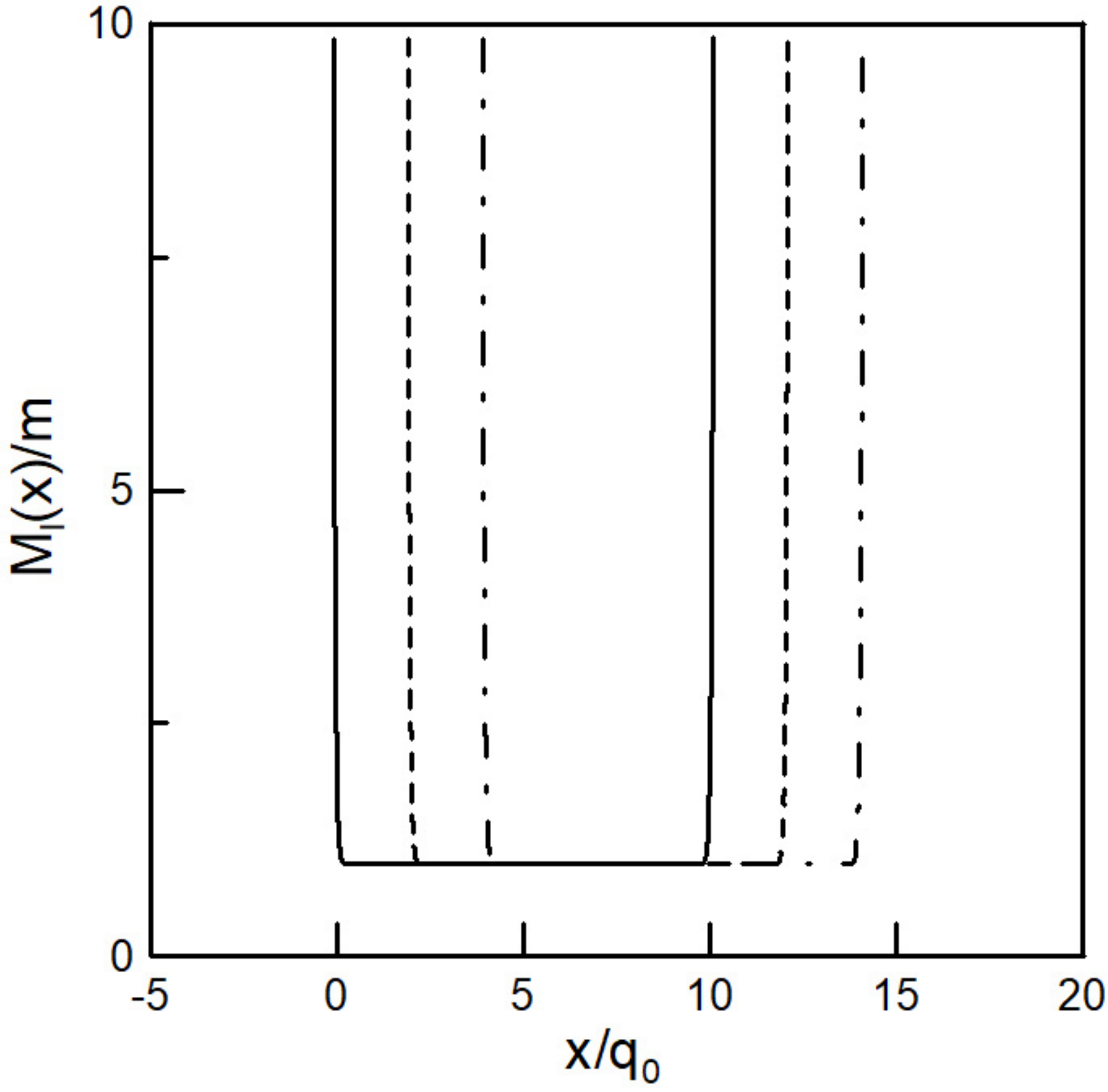}
\end{center}
\caption{The behavior of the quantum PDM $M_\ell(x)$ in units of the classical $m$. We set $\ell = 0.1 q_0$.
The solid, dashed and dot-dashed lines represent the results of $a/q_0 =0,2$ and $4$, respectively.}
\label{fig:Mass}
\end{figure}

Applying the CS quantization \eqref{quantchif}, the corresponding Schr\"{o}dinger equation is given by 
\begin{eqnarray}
\ii \hbar \partial_t \phi 
= \hat{A}_{H_\chi}  \phi\,,
\end{eqnarray}
where
\begin{eqnarray}
\label{order1}
\hat{A}_{H_\chi} 
&=& \left[
\frac{1}{4} \left\{ \frac{1}{M_\ell (\hq)}, \hp^2 \right\} + V^+ (\hq) 
\right]
=
\left[
-\frac{\hbar^2}{4} \left\{ \frac{1}{M_\ell (x)}, \partial^2_x \right\} + V^+ (x) 
\right]\,, 
\end{eqnarray}
or equivalently 
\begin{eqnarray}
\label{order2}
\hat{A}_{H_\chi} 
&=& \left[
\frac{1}{2} \hp\frac{1}{M_\ell(\hq)} \hp   +V^- (\hq) 
\right]=\left[
-\frac{\hbar^2}{2} \partial_x \frac{1}{M_\ell(x)} \partial_x   +V^- (x) 
\right] \, .
\end{eqnarray}
Note that the symmetric quantum operators yielded by the Weyl-Heisenberg integral quantization, with coherent states or with more general POVM,  of Galilean shadow invariant Hamiltonians of the  general form $H= a(q) p^2 + b(q) p + c(q)$,  are given in \cite{becugaro17,gazeau18}.

In \eqref{order1} and \eqref{order2} we have introduced three multiplication operators, namely, the PDM induced by the quantization, 
\begin{eqnarray}
M_\ell(x) = \frac{m}{B_{\ell}(x,a,b)}\, , 
\end{eqnarray}
which should be compared to the classical one, (\ref{posdepm}),
and the two potential terms  defined by 
\begin{equation}
\begin{split}
V^{\pm} (x) 
&= 
\frac{\hbar^2}{4\ell^2 M_\ell(x)}
\left( 
1
\pm \frac{\ell^2 B_{\ell}^{(2)}(x,a,b)}{2B_{\ell}(x,a,b)}
\right)\\&= \frac{1}{4}\frac{\hbar^2}{\ell^2}\left(\frac{B_{\ell}(x,a,b)}{m}\pm \frac{1}{m \sqrt{\pi}}\left(\frac{x-a}{\ell}e^{-\left(\frac{x-a}{\ell}\right)^2}-\frac{x-b}{\ell}e^{-\left(\frac{x-b}{\ell}\right)^2}\right)\right)\,.
\end{split}
\end{equation}
These two different potentials come from the choice between the two symmetric orderings Eq.\ \eqref{order1} 
and Eq.\ \eqref{order2}. One notices that  these potentials vanish at the classical limit, as expected.

\begin{figure}[t]
 \begin{minipage}[b]{0.4\linewidth}
  \centering
  \includegraphics[keepaspectratio, scale=0.2]{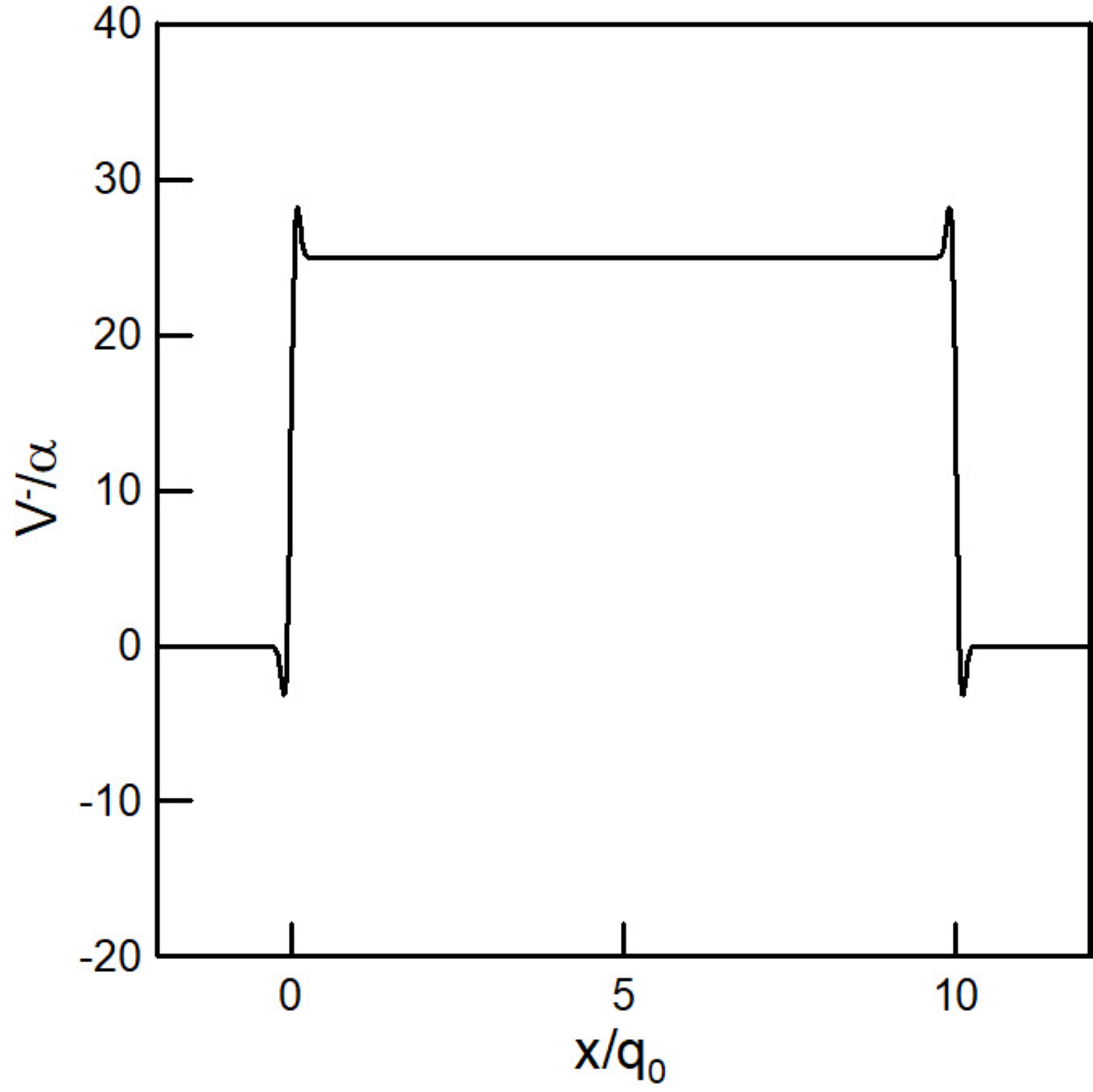}
 \subcaption{$\ell$ = 0.1$q_0$ }
 \end{minipage}
 \begin{minipage}[b]{0.4\linewidth}
  \centering
  \includegraphics[keepaspectratio, scale=0.2]{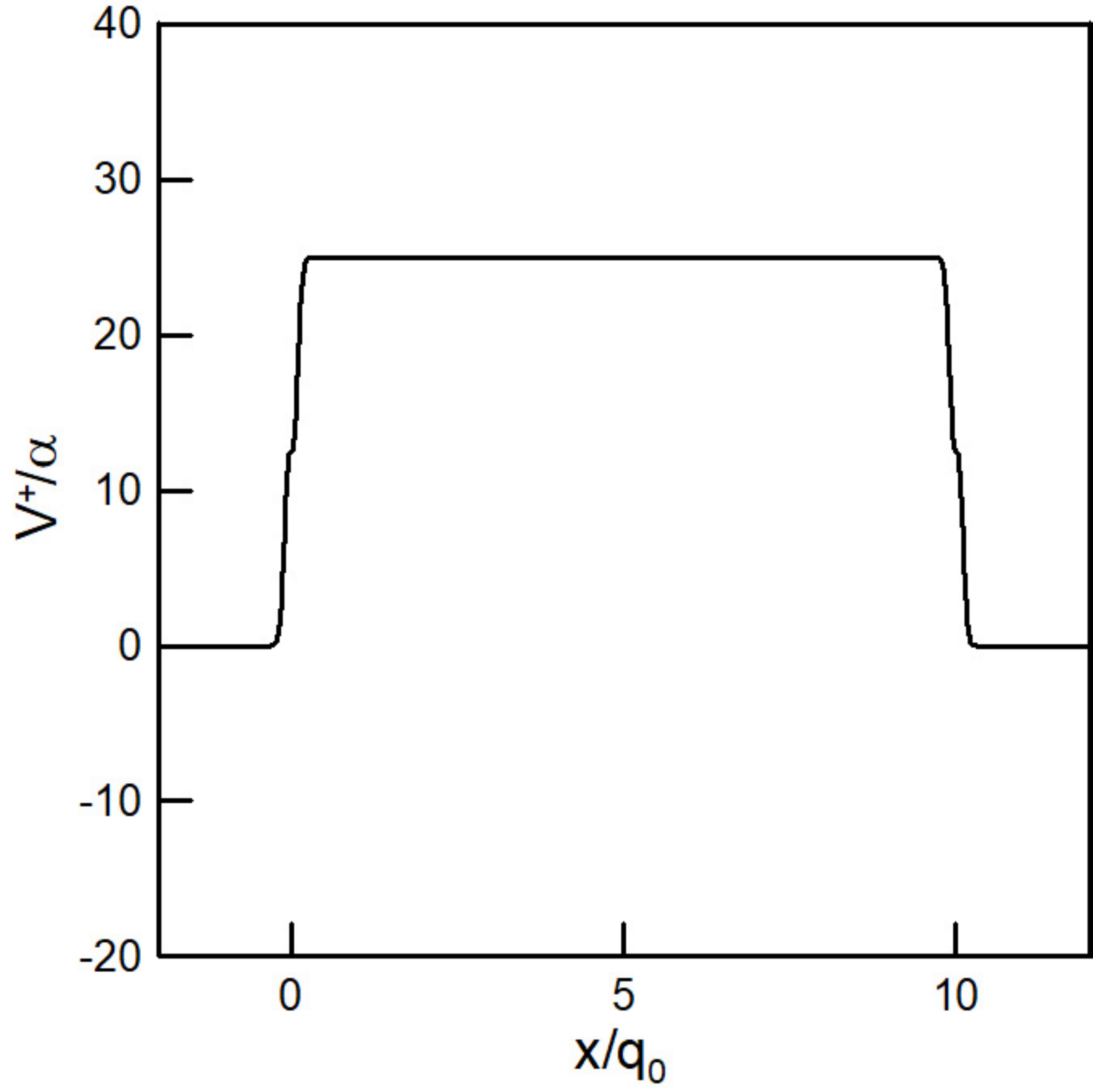}
\subcaption{$\ell$ = 0.1$q_0$ }
 \end{minipage}
\\
 \begin{minipage}[b]{0.4\linewidth}
  \centering
  \includegraphics[keepaspectratio, scale=0.2]{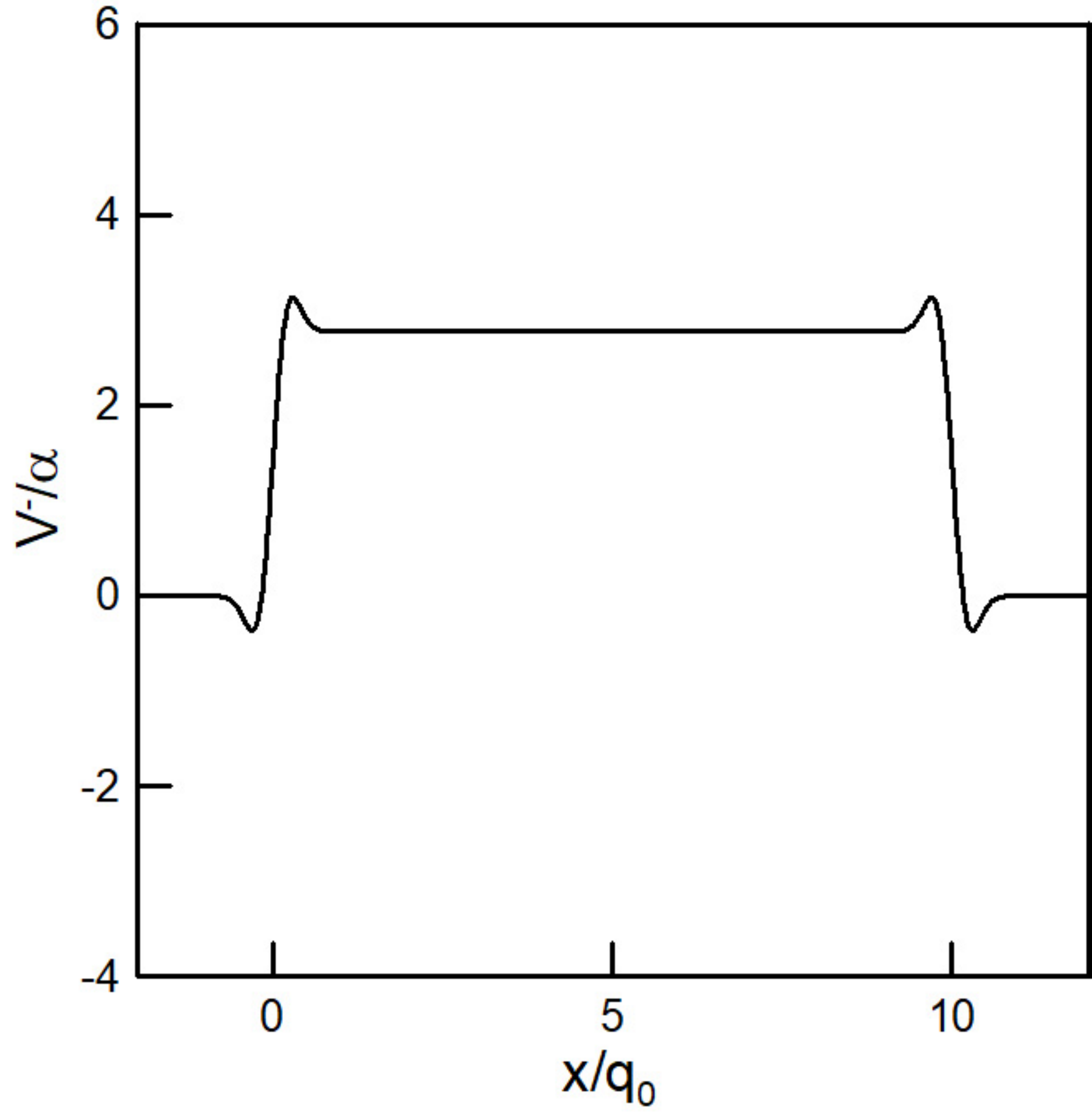}
 \subcaption{$\ell$ = 0.3$q_0$ }
 \end{minipage}
 \begin{minipage}[b]{0.4\linewidth}
  \centering
  \includegraphics[keepaspectratio, scale=0.2]{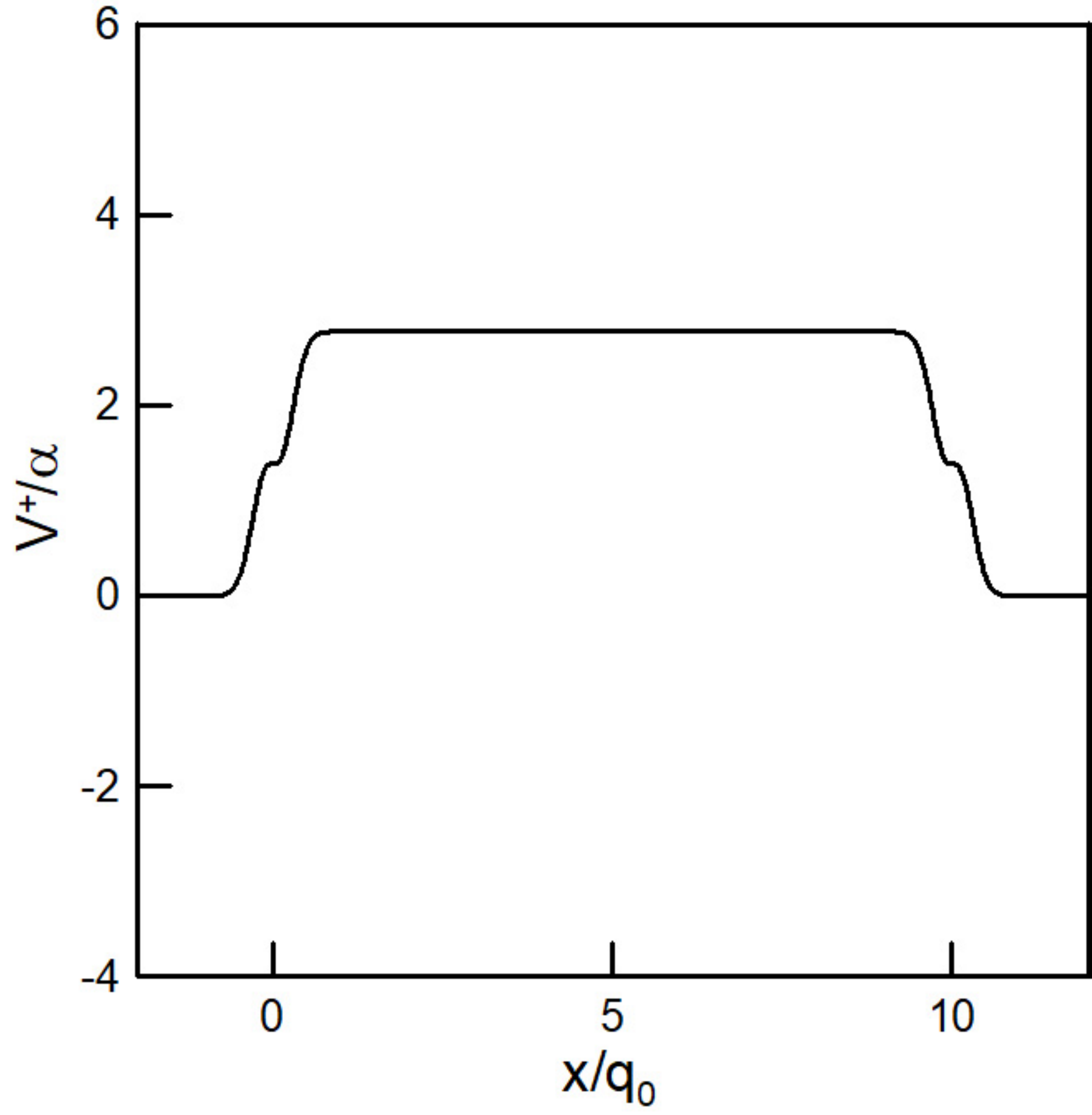}
 \subcaption{$\ell$ = 0.3$q_0$ }
 \end{minipage}\\
 \begin{minipage}[b]{0.4\linewidth}
  \centering
  \includegraphics[keepaspectratio, scale=0.2]{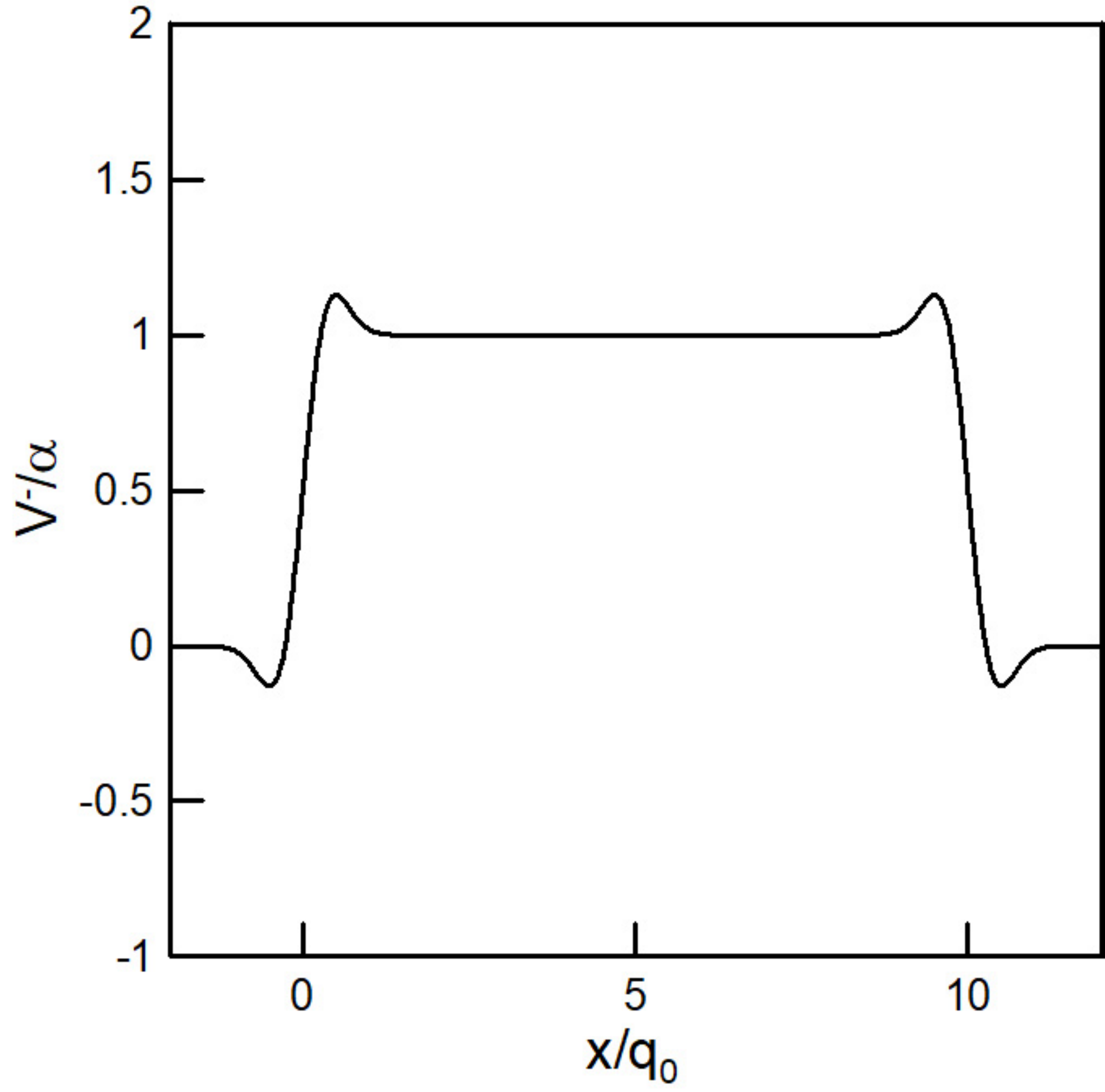}
 \subcaption{$\ell$ = 0.5$q_0$ }
 \end{minipage}
 \begin{minipage}[b]{0.4\linewidth}
  \centering
  \includegraphics[keepaspectratio, scale=0.2]{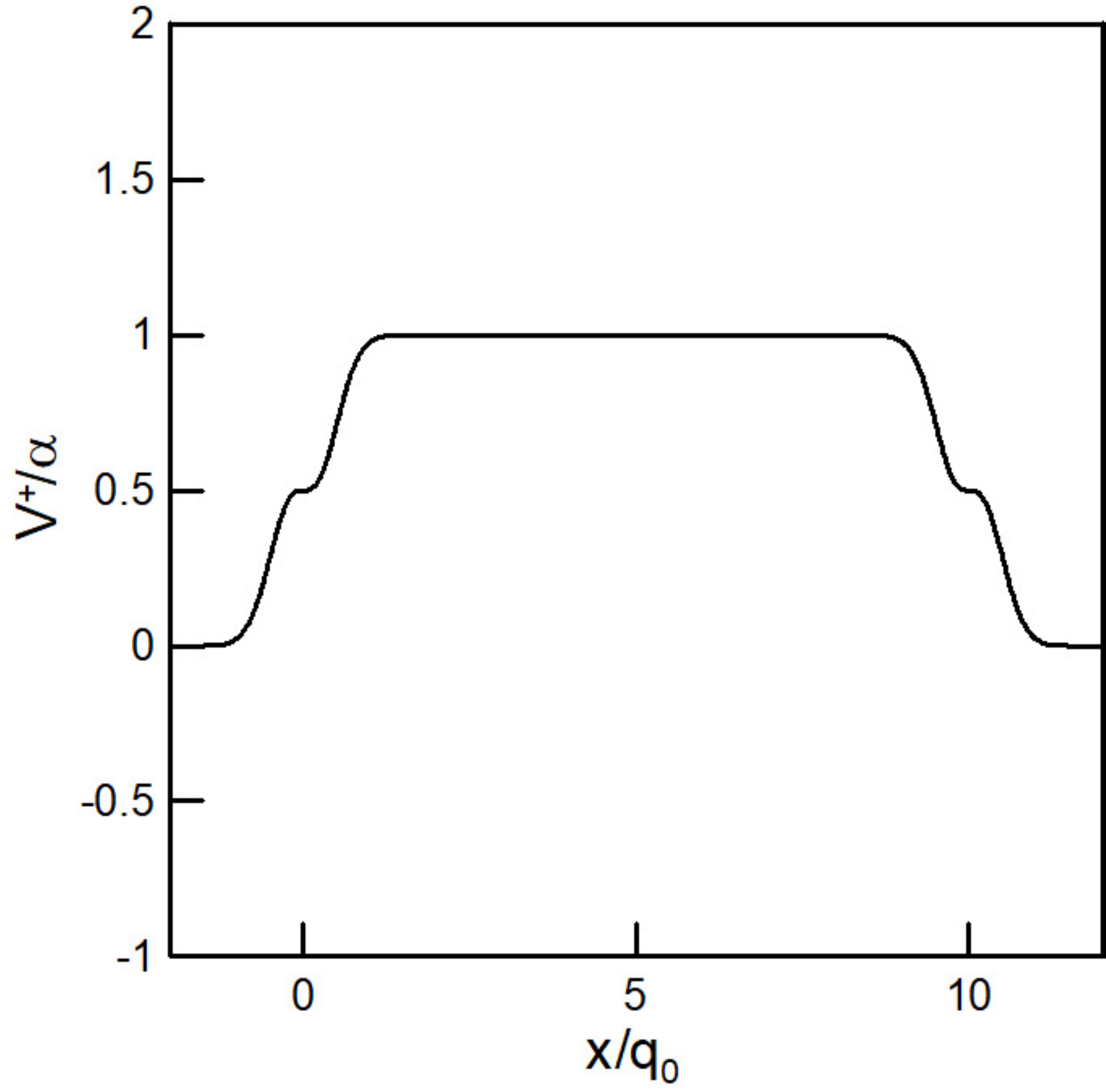}
 \subcaption{$\ell$ = 0.5$q_0$ }
 \end{minipage}
\caption{The behaviors of the functions $V^-(x)$ (left panel) and $V^+(x)$ (right panel) in units $\alpha = \hbar^2/(mq_0^2)$. 
We set $\ell = 0.1q_0$ (top) $0.3q_0$ (middle) and $0.5q_0$ (bottom). 
The parameter $a$ is set to be $0$. 
}
\label{fig:pot}
\end{figure}

In  previous considerations on quantum PDM Hamiltonians, e.g., \cite{leblond95}, it is expected that the free Hamiltonian can be expressed without introducing potential terms as
\begin{eqnarray}
\label{eqn:h-order}
\hat{H} = \frac{1}{4} (M^\alpha(x)\hat{p}M^\beta(x)\hat{p}M^\gamma(x) + M^\gamma(x)\hat{p}M^\beta(x)\hat{p}M^\alpha(x)  ), \ \ \  (\alpha+\beta + \gamma = -1),
\end{eqnarray}
if $\alpha$, $\beta$ and $\gamma$ are chosen appropriately. In our approach,  the appearance of a potential term cannot be avoided, whatever the ordering choice.

The behavior of the PDM operator is shown in Fig.\ \ref{fig:Mass}. 
We set $\ell = 0.1q_0$ and 
the solid, dashed and dot-dashed lines represent the results of $a/q_0 =0,2$ and $4$, respectively.
One can see that the mass increases rapidly to prevent the particle from escaping from the classical domain  $a<x<b$.
It should be noted that, in the classical limit, this function goes to the singular PDM introduced in Eq.\ \eqref{posdepm}
\begin{eqnarray}
M_\ell(x) \underset{\ell \to 0}{\to} m_{1/\chi}(x)\,.
\end{eqnarray}

Similarly, the potentials $V^-(x)$ and $V^+(x)$ are, respectively, shown on the left 
and right panels of Fig.\ \ref{fig:pot} for different values of $\ell$
in units of $\alpha = \hbar^2/(mq^2_0)$. 
We set $\ell = 0.1q_0$ (top) $0.3q_0$ (middle) and $0.5q_0$ (bottom). 
The parameter $a$ is set to be $0$.
Both behaviors are very close: $V^{\pm}(x)$ is almost constant and 
shows non-trivial changes 
near the endpoints, which are enhanced as $\ell$ decreases.
This is due to the behavior of the PDM itself. 
We note that $V^+(x)$ and $V^-(x)$ vanish in the classical limit, as expected.

\section{Semi-classical Hamiltonian mechanics and its classical limit}
\label{semclass}
If the PDM system is quantized appropriately through our approach, 
its semi-classical behavior should be very close to the corresponding classical dynamics. 
In other words, the examination of the semi-classical portrait is important to confirm the consistency of our approach. 
Let us determine the dynamics induced by the semi-classical phase-space portrait of the quantum Hamiltonian $ \hat{A}_{H_\chi}$. The latter is given by
the smooth function
\begin{eqnarray}
\check H_\chi(q,p)
=\langle q,p | \hat{A}_{H_\chi} | q,p\rangle = 
B_{\sqrt{2}\ell}(q,a,b) \left( 
\frac{p^2}{2m} + \frac{\hbar^2}{2m\ell^2}\right)
\,. \label{eqn:semi-tra-p}
\end{eqnarray}
We note that this semi-classical Hamiltonian is a PDM one of the type \eqref{Hvar},
\begin{equation}
\label{Hchivar}
\check H_\chi(q,p)= \frac{p^2}{2M_{\sqrt{2}\ell}(q)} + V_{\ell}(q)\,,
\end{equation}
where
\begin{eqnarray}
V_{\ell}(q):= \frac{\hbar^2}{2m\ell^2} \, B_{\sqrt{2}\ell}(q,a,b) = \frac{\hbar^2}{2 M_{\sqrt{2}\ell}(q)\,\ell^2}\,.
\end{eqnarray} 
The resulting semi-classical dynamics is then obtained from the general formulae given in the previous section.
\begin{eqnarray}
\label{dotqsc}
\dot{q} 
&=& \{ q, \check H_\chi \}_{PB} 
= \frac{p}{M_{\sqrt{2}\ell}(q)}, \\
\label{dotpsc} \dot{p}
&=& \{ p, \check H_\chi  \}_{PB} = - \frac{\ud V_\ell}{\ud q}(q) + \frac{p^2}{2M_{\sqrt{2}\ell}^2(q)}\frac{\ud M_{\sqrt{2}\ell}}{\ud q}(q)\, .  
\end{eqnarray}
The force exerted on the particle is, according to our definition \eqref{classforce},
\begin{equation}
\label{semclassforce}
F_\ell(q,p)= M_{\sqrt{2}\ell}(q) \ddot{q}  
= - \frac{\ud V_\ell}{\ud q}(q) - \frac{p^2}{2M^2_{\sqrt{2\ell}(q)}} \frac{\ud M_{\sqrt{2}\ell}}{\ud q}(q)
= B^{(1)}_{\sqrt{2}\ell}(q,a,b) \left( 
\frac{p^2}{2m} - \frac{\hbar^2}{2m\ell^2}\right)\, .
\end{equation}
One can easily see that, 
while the semi-classical equations qualitatively reproduces the corresponding classical ones, 
this semi-classical behavior depends on the momentum of the particle due to the quantum effect: 
the semi-classical force gives rise to the effect of confinement 
only for the particle which has the magnitude of the momentum $\vert p\vert  > \hbar/\ell$, a lower bound which can be made arbitrary small at large $\ell$.  
The behavior is shown in Fig.\ \ref{scforce}.
The left and right panels show the results 
of $p=0$ and $p=20\hbar/q_0$, respectively. The solid, dot-dashed and dashed lines represent $\ell = 0.1$, $0.3$ and $0.5$, respectively.
The semi-classical force for the  right panel shows the confinement. It is interesting to notice the  critical absolute value of the momentum, $\vert p\vert_c= \hbar/\ell$ for which there is no force at the boundary. According to \eqref{classlim} this ratio has to be considered as a small quantity.

\begin{figure}
\begin{center}
\includegraphics[width=2in]{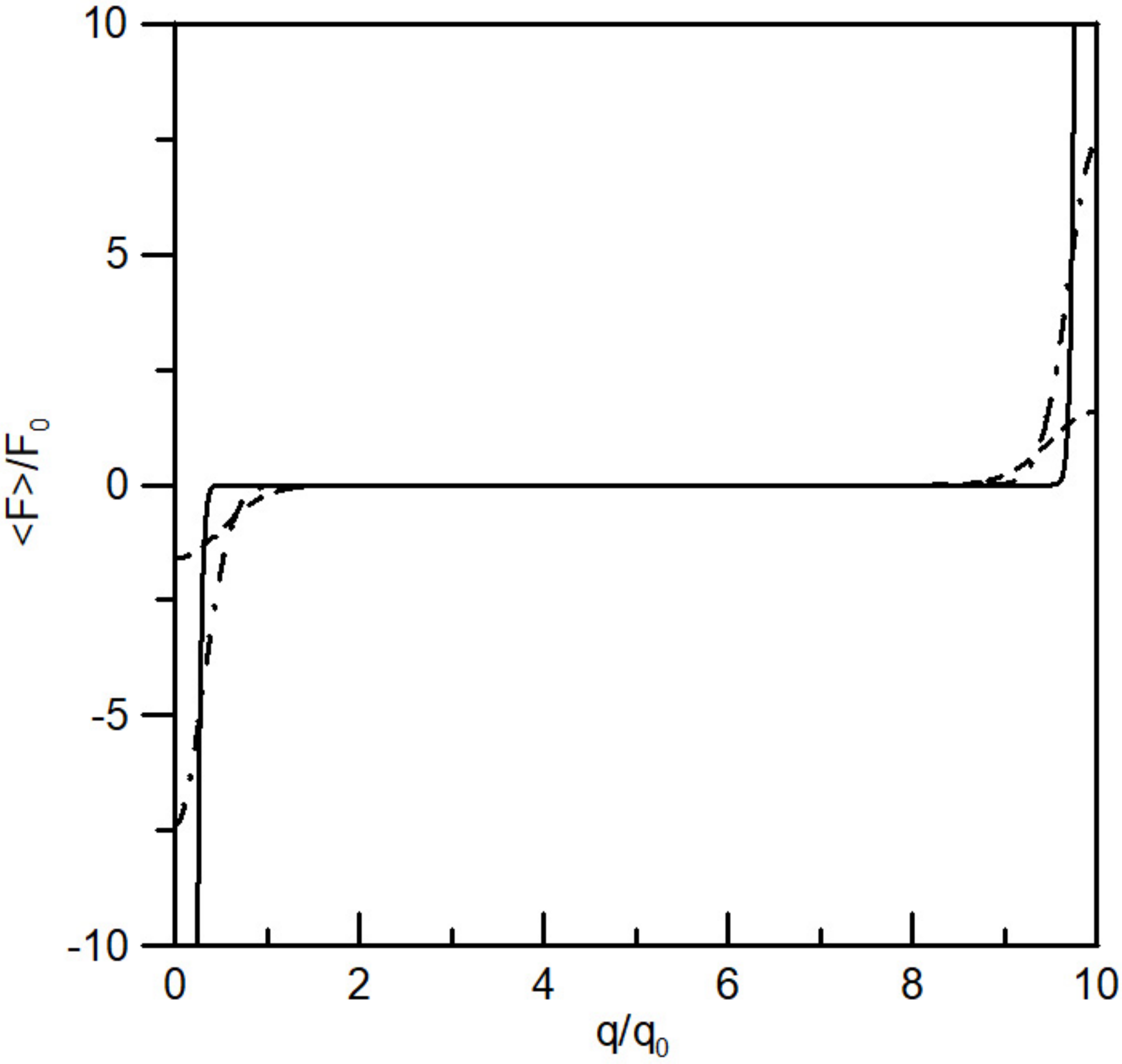}
\includegraphics[width=2in]{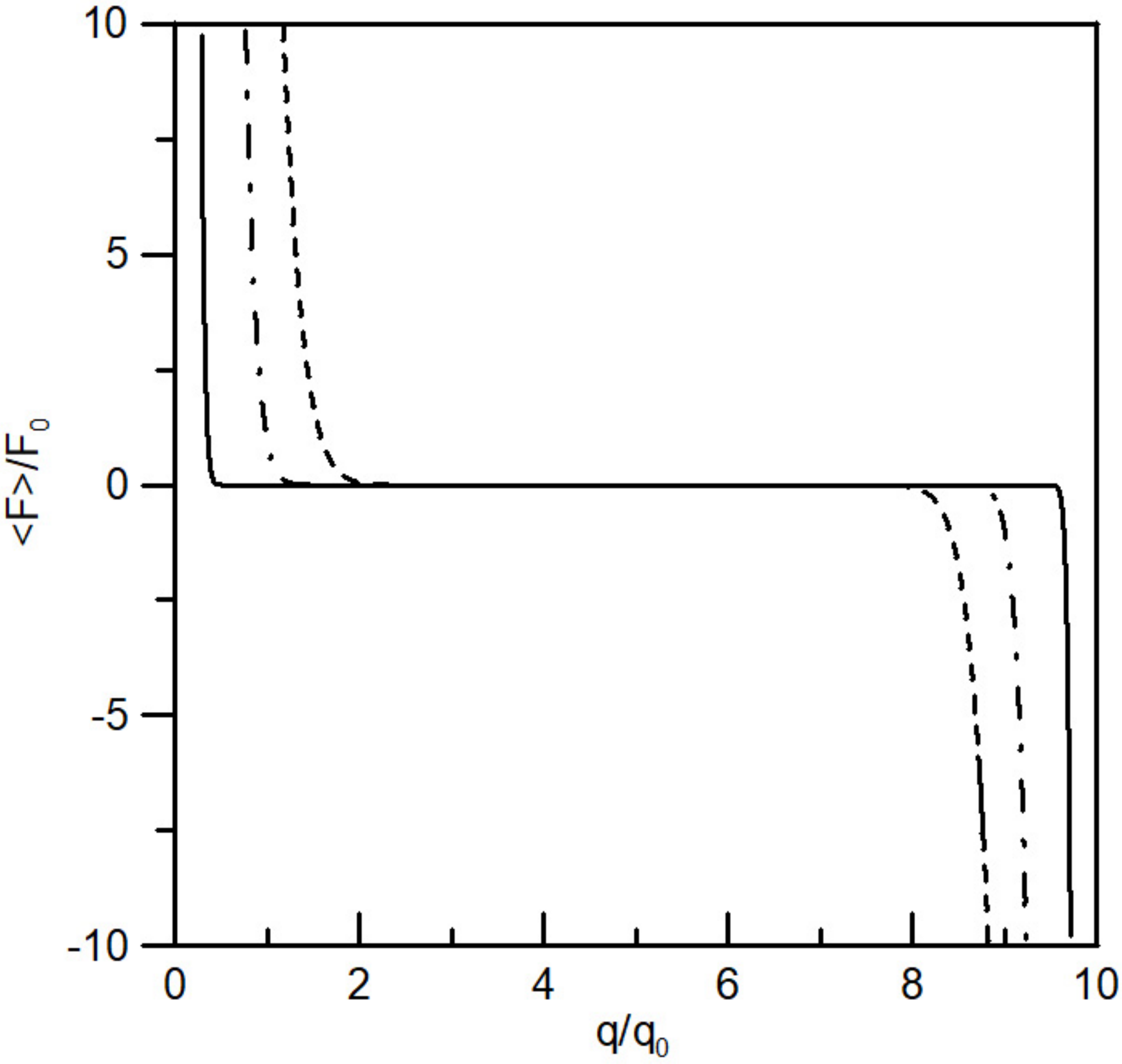}
\caption{The position dependence of the semi-classical forces. The left and right panels show the results 
of $p=0$ and $p=20\hbar/q_0$, respectively. The solid, dot-dashed and dashed lines represent $\ell = 0.1$, $0.3$ and $0.5$, respectively. We set $F_0 = \hbar^2/(2m q^2_0)$. The semi-classical force for the right panel shows the confinement.}
\label{scforce}
\end{center}
\end{figure}

In the classical limit \eqref{classlim}, the above equations become
 \begin{eqnarray}
\label{dotqc}
\dot{q}_{\mathrm{class}}  
&=& \frac{p_{\mathrm{class}}}{m}, \\
\label{dotpc} \dot{p}_{\mathrm{class}} 
&=&   -\frac{p^2_{\mathrm{class}} }{2m}\left(\delta_a(q_{\mathrm{class}})-\delta_b(q_{\mathrm{class}})\right)\, .  
\end{eqnarray}
\begin{equation}
\label{classforce2}
F_{\mathrm{class}} 
= \frac{p^2_{\mathrm{class}} }{2m}\left( \delta_a(q_{\mathrm{class}})-\delta_b(q_{\mathrm{class}})\right)\, .
\end{equation}
They should be viewed as a consistent solution to the dynamics ruled by the truncated Hamiltonian \eqref{Hvarchi} with $q_{\mathrm{class}} \equiv q_\chi$ and $p_{\mathrm{class}} \equiv p_\chi$. 
The above equation for the classical force is precisely what we can expect of the infinite repulsive action from the walls in the case of the infinite square  well. 
In Fig.\ \ref{scforce} is shown the behavior of the semi-classical force \eqref{semclassforce} as a smooth version of the singular \eqref{classforce2}.

\section{Conclusion}
\label{conclu}

We have generalized the Weyl-Heisenberg covariant integral quantization of a one-dimensional classical system constrained to lie within a bounded or semi-bounded geometry.  
We have  found that our quantization regularizes the discontinuous classical position-dependent mass and furthermore introduces an extra potential.
To our best knowledge, the possibility of the modification of the form of the classical PDM by quantization has been overlooked in existing studies 
thus far and this is one of the important results yielded by our  approach. 
Moreover, it has been considered that the potential term induced by the quantization of the PDM systems 
is attributed to the ambiguity for the ordering of operators and, if we choose it appropriately, such a potential term is 
absorbed by the kinetic operator in the quantum Hamiltonian as is shown in Eq.\ (\ref{eqn:h-order}). We however showed that the integral quantization of PDM, 
which is free of the ordering problem of operators, leads to the new potential term and then the Hamiltonian operator cannot be expressed in the form of  Eq.\ (\ref{eqn:h-order}).

To justify our approach, we discussed the semi-classical portrait of the derived quantum dynamics.
The semi-classical behavior describes the bounded motion under the geometric constraint and 
the equation of motion reproduces the corresponding classical equation besides the corrections induced by the quantum effects.
Our approach is consistent in this sense.

We also examined  the question of (essential) self-adjointness of quantum observables like momentum and Hamiltonian for the confined free classical particle. 
Based on our results we have obtained for the interval, we  expect that the appearance on the quantum level of smooth PDM and semi-confinement potentials could get rid of the ambiguity of imposing boundary conditions to the wave functions in solving the Schr\"{o}dinger equation. 
These problems deserve to be seriously considered in future investigations. 

In our scheme, some characteristic length besides the Planck constant, $\ell$, is introduced, This regularization parameter $\ell$ can be adjusted in order to fit in an optimal way experimental models for nanostructures like $1$d quantum dots. The extension of the approach  to $2$d or $3$d is straightforward.  These new degrees of freedom opens channels for future investigation concerning more realistic models, like a manifold $E$ embedded into $\R^d$, for which the coherent states to be used would be $\otimes_{i=1}^d |q_i,p_i\rg$. Hence, we could compare our results with previous ones obtained through different approaches, e.g. \cite{costa81,schujaff03}, particularly those concerning forces resulting from boundary surfaces.   Note that a similar approach has allowed to regularize the so-called Bianchi $9$ potential in cosmology, as is shown in the  recent paper \cite{berczugama18}.

Another appealing direction in the application of our approach would consist in considering systems moving in punctured geometries, like the simplest $\R\setminus\{0\}$, or more generally $\R^d\setminus E$, where $E$ is a certain subset, for instance a lower-dimensional manifold or a discrete subset. Therefore, instead of quantizing a classical observable $f(q,p)$, 
we could quantize $f(q,p)(1 - \delta_E(q))$, and analyze the resulting quantum dynamics 
in a  way similar to the present work.

\hspace{2cm}

T.\ K.\ acknowledges the financial support by CNPq (307516/2015-6,303468/2018-1). 
A part of the work was developed under the project INCT-FNA Proc.\ No.\ 464898/2014-5. J.\ P.\ G. thanks CBPF for hospitality.

\end{document}